## An effective interest rate cap: a clarification

Mikhail V. Sokolov[a,b,c,d]

[a] European University at St. Petersburg, 6/1A Gagarinskaya st., St. Petersburg, 191187, Russia
[b] Centre for Econometrics and Business Analytics (CEBA). St. Petersburg State University, 7/9 Universitetskaya nab., St. Petersburg, 199034, Russia
[c] Institute for Regional Economic Studies RAS, 38 Serpukhovskaya st., St. Petersburg, 190013, Russia
[d] HSE University, 16 Soyuza Pechatnikov st., St. Petersburg, 190121, Russia

**Abstract** Many countries impose regulatory restrictions on lending rates known as interest rate caps. In most cases, these restrictions apply to the effective (rather than nominal) interest rate, a measure which incorporates all commissions and fees associated with a loan. Because the effective interest rate is the internal rate of return (IRR) of the loan's cash flow stream, this regulatory rule becomes ambiguous for loans that do not have a conventional IRR. This paper resolves this ambiguity. We begin by clarifying the concept of IRR. We axiomatize the conventional definition of IRR (as a unique root of the IRR polynomial) and demonstrate that any extension to a larger domain necessarily violates a natural axiom. Building on this result, we show that there is a unique extension of the interest rate cap to all loans consistent with a set of economically meaningful axioms. The rule we characterize takes the form of a net present value test. This result is general, and applies to any setting where one wishes to extend an IRR-based threshold rule to arbitrary cash flows. Applications include lending and deposit rates regulation, investment screening, and capital budgeting, where the standard decision rule accepts a project if its IRR exceeds the hurdle rate.

---

E-mail address: mvsokolov@eu.spb.ru
I am grateful to James Kolari, Aleksei Kondratev, Leonid Mitrofanov, Alexander Nesterov, Ekaterina Polyakova, Kirill Romanyuk, and Dmitriy Skougarevskiy for helpful comments that greatly improved the paper.
This work was supported by HSE University (Basic Research Program).

## 1. Introduction

At least 76 countries around the world impose regulatory restrictions on lending rates in the form of interest rate caps (or ceilings) (Maimbo and Gallegos 2014; Ferrari et al. 2018). There are several economic and political rationales for such regulation. One is to provide support to specific industries or sectors of the economy, where market failures exist. Those market failures usually result from information asymmetries and the inability of financial institutions to differentiate between risky and safe clients, from adverse selection, and from moral hazard (Stiglitz and Weiss 1981). Another common rationale is to protect consumers from usury and prevent the most vulnerable and financially less literate segments of the society from predatory lending. Those rationales are, however, disputed by some scholars and policymakers who argue that an interest rate cap is an inefficient and distorted tool for lowering rates, which has substantial unintended side-effects (see Ferrari et al. 2018 for a review of this evidence). Critics of interest rate caps also sustain that this regulatory measure limits consumer choice and thus harms the consumers they are supposed to help (see Miller and Black 2016 for a survey of this literature).

In this paper, we leave aside the debate on whether an interest rate cap is an appropriate tool of interest rate control, but concentrate on how it should be implemented if a policymaker decides to employ it. In what follows, we, rather loosely, refer to the formulation of this regulatory measure as a usury law since it is the most common legal instrument for implementing interest rate caps (Maimbo and Gallegos 2014). In most cases, the effective (rather than nominal) interest rate is restricted, which includes all fees, commissions, and other expenses associated with a loan (Maimbo and Gallegos 2014). Typically, the generic wording of such a usury law is ambiguous in three respects.

A. Recall that the effective interest rate is the internal rate of return (IRR) of the cash flow stream associated with a loan. The investment appraisal literature provides several nonequivalent concepts of IRR.[1] A usury law, therefore, should specify which one is used.

B. Most definitions of IRR are partial in the sense that there are cash flow streams that have no IRR. How should a usury law be interpreted for those loans whose cash flow streams have no IRR?

C. When loan advances and repayments alternate in time, the respective roles of the borrower and lender can be blurred. In this case, it is unclear which of the parties of the loan contract a usury law should be addressed to and, if the law treats the contract as illegal, which party should be brought to justice.

Ambiguity A seems purely technical and can be resolved in a normative way. However, it affects ambiguity B; so they should be clarified cooperatively. According to the prevalent definition, the IRR of a cash flow is defined as the discount rate at which the cash flow has zero net present value. Ambiguity B arises if the net present value of the cash flow as a function of the discount rate—the IRR polynomial—has more than one root or no roots at all. This situation cannot be ignored for at least two reasons. First, it occurs with regular frequency in particular industries or types of financial products. For instance, for some types of loans (e.g., consumer loans), it is usual to charge lender fees (e.g., an application fee, origination fee, processing fee, or monthly service fee) to cover costs associated with underwriting and processing a loan. Such fees are most common in a mortgage loan, which typically includes several ad hoc fees in addition to the monthly interest. A usury law treats such fees as a part of the cash flow associated with the loan. In this case, the resulting cash flow stream has more than one change of sign, which usually results in more than one root of the IRR polynomial. In particular, the presence of an application fee—probably the most common type of lender fee—charged before a loan is processed, *necessarily* results in a cash flow stream that has no IRR in the conventional sense (more precisely, it is not true that the IRR polynomial has exactly one sign change), and therefore, the loan cannot be evaluated by a typical usury law. As another example, some types of loans are accompanied with regular frequency by a refund. For instance, in some countries, borrowers of consumer loans who repay their debts early are eligible for a refund of the insurance premium for all insured risks, provided that no insured event has occurred. Again, the cash flow stream of such a loan has no IRR and therefore cannot be evaluated by a typical usury law. However, following the spirit of a usury law, if the law authorizes a loan, then it must also authorize this loan accompanied by a refund as the refund makes the loan more attractive to the borrower.

[1] The most widespread definition of IRR is a root of the IRR polynomial, provided that the root exists and is unique. However, some authors argue that the root uniqueness condition is not sufficient to be relevant for decision making. For instance, Herbst (1978) asserts that IRR is a proper measure of return on investment just for conventional investments that have only one change of sign in their net cash flow streams. Gronchi (1986) and Promislow (2015, Section 2.12) argue that IRR is meaningful only for so called pure investment and borrowing streams, introduced in Teichroew et al. (1965). Some authors require IRR to be a simple root of the IRR polynomial (Norstrøm 1972; Vilensky and Smolyak 1999), the condition guaranteeing continuity of IRR as a function of cash flow stream. In contrast, multiple generalizations of the common definition of IRR are proposed: Arrow and Levhari (1969), Cantor and Lippman (1983), Promislow and Spring (1996), to mention just a few. The balance function approach (Teichroew et al. 1965; Spring 2012), the proposal of Hazen (2003), the relevant IRR (Hartman and Schafrick 2004), the average IRR (Magni 2010, 2016), and the selective IRR (Weber 2014) provide generalizations of IRR conditional on exogenously given reinvestment rate, cost of capital, or capital stream.

Second, a lender who knew that the usury law did not deal adequately with loans that differed from the standard pattern could deliberately create such a situation to get around the law. This can easily be implemented as each cash flow stream possessing a unique IRR can, by an arbitrary small perturbation (such as receiving a monetary unit before the initial outlay or paying a monetary unit after the final inflow), be transformed into a cash flow stream that has no unique IRR.

Let us illustrate ambiguity B by examples from Canada's anti-usury legislation. Canada's Criminal Code prohibits lending money at an effective annual rate exceeding $60\%$, where the effective rate includes all charges and expenses paid or payable for the advancing of credit, with only a few exceptions.[2] Given that the effective rate is the IRR of the cash flow stream associated with the loan, how should this provision be interpreted for loans whose cash flows have no IRR? To illustrate, first consider a 1-year consumer loan of 100 Canadian dollars at an interest rate of 70% payable at maturity. Clearly, this loan, which we denote by $x$, has an effective interest rate of 70% and, hence, the Code considers it as usurious (illegal). Now consider the same loan accompanied with an application fee of 1% of the loan amount paid 1 day prior. Assuming that time is measured in years, the lender's cash flow stream associated with this loan, which we denote by $y$, is 1 dollar at time $0$ (today), $-100$ dollars at time $1/365$ (tomorrow), and 170 dollars at time $366/365$ (a year and a day). The IRR polynomial associated with $y$, $s \mapsto 1-100(1+s)^{-1/365}+170(1+s)^{-366/365}$, has in $(-1,+\infty)$ two real roots. So the effective interest rate for $y$ is undefined; the Code is inapplicable and therefore considers $y$ as legal. This is counterintuitive as $y$ is even more favorable to the lender than the usurious loan $x$. This example shows that the naive approach of "if the effective interest rate is undefined, the loan is legal" fails.

[2] Criminal Code, Revised Statutes of Canada 1985, c. C-46, Section 347.

Consider again the loan $x$, which the Code renders illegal. To evade the law, an unscrupulous lender can accompany $x$ with a small refund of $c$ dollars the day after the loan repayment (note that loans with a refund—usually subject to certain conditions being met—occur with regular frequency in particular types of financial products; see Section 5 (part 2º) for real-world examples). For any $c>0$, the IRR of the resulting loan, which we denote by $z_c$, is undefined as it is not true that the IRR polynomial $s \mapsto -100+170(1+s)^{-1}-c(1+s)^{-366/365}$ has in $(-1,+\infty)$ exactly one sign change. So $z_c$ has no effective rate; it cannot be evaluated by the Code and therefore is considered as legal. At least for small $c>0$, this conclusion is counterintuitive as $z_c$ is a minor perturbation of the 'highly usurious' loan $x$. These as well as several other potential issues of Canada's anti-usury legislation are documented in the law and actuarial literature (Promislow 1997; Waldron 2002; Karp 2003; ULCC 2008).

Finally, ambiguity C arises when loan advances and repayments alternate in time. In this case, the respective roles of the borrower and lender can be blurred. This situation can occur even if the loan has a unique IRR (Herbst 1978; Gronchi 1986). It more often arises for loans that come from non-traditional money lenders like individuals or companies instead of banks, since those loan agreements, in contrast to bank loan agreements, are non-standardized. To illustrate the problem, assume that X and Y sign a contract, according to which X receives from Y the transaction $(1,-3,2)$, i.e., X receives 1 dollar at time 0, pays 3 dollars at time 1, and receives 2 dollars at time 2. One would treat this contract as an interest-free loan, since, in sum, the parties transfer each other 3 dollars. However, it is unclear who is the lender and who is the borrower in the contract and, thus, which of the parties of the contract a usury law should be addressed to. For instance, one might treat Y (resp. X) as the lender, because the transaction for Y (resp. X) starts (resp. ends) with an outflow (resp. inflow). Actually, neither of the parties is the pure lender/borrower as the transaction involves both borrowing and lending operations at the same interest rate, e.g., $(1,-2,0)$ and $(0,-1,2)$. If a usury law treats this contract as usurious, it is unclear which party should be brought to justice.

Employing the definition of IRR in the sense of Gronchi (1986) and Promislow (2015, Section 2.12) (that is, the restriction of the conventional definition of IRR as a unique root of the IRR polynomial to the set of so called pure investments introduced in Teichroew et al. (1965, pp. 155–156)), problem B was studied in detail in Promislow (1997). The author examined it from an axiomatic viewpoint and proved an impossibility result, showing that under a certain natural set of axioms, there is no general solution to this problem. By relaxing the requirement that all loans be classified, various solutions were obtained. Problem B is also closely related to the question of whether the concept of IRR can be extended to the set of all cash flows. Indeed, if there is such an extension, then its corresponding lower (resp. strict upper) contour set is precisely the set of legal (resp. illegal) loans. Though the investment appraisal literature provides a variety of such extensions, as shown in Promislow (1997), Vilensky and Smolyak (1999), and Sokolov (2024), any extension of the conventional definition of IRR to the set of all cash flows necessarily violates a set of reasonable axioms.

This paper aims to resolve ambiguities A–C. We start by defining IRR via an axiomatic approach. Our axiomatization is along the lines of Vilensky and Smolyak (1999). We show that the conventional definition of IRR (as well as its restriction to a proper subset) as a unique root of the IRR polynomial is the only one consistent with the two natural axioms (Proposition 2). Moreover, the IRR defined this way cannot be extended to a larger set of cash flows. This result shows that the concept of IRR that needs to be specified to eliminate ambiguity A must be either the conventional one or its restriction to a proper subset.

Given this result and using an axiomatic approach, we show how to extend the generic statement of a usury law (which is currently only applicable to loans possessing IRR) to all loans and, thus, resolve ambiguity B. In particular, we prove that there is a unique extension consistent with the conventional definition of IRR (Proposition 3). This extension takes the form of a net present value (NPV) test: given a maximum allowable effective interest rate $r$, a loan is usurious if and only if the lender's cash flow has positive NPV at some discount rate $s \geq r$. For instance, turning back to the examples from Canada's anti-usury legislation mentioned above, we have $r = 60\%$, and the loans $x$ and $y$ are usurious (as their NPVs are positive at the rate 60%) and the loan $z_c$ is usurious if and only if $c < 10.013$ (as its IRR polynomial is nonpositive on $[60\%, +\infty)$ if and only if $c \geq 10.013$). More crucially, we reveal the geometric structure of the set of legal loans in the general case: given a maximum allowable effective interest rate $r$, we show that, irrelevantly of the definition of IRR chosen, the set of legal loans is the polar cone of a collection of NPV functionals whose discount functions have instantaneous discount rates of at least $r$ (Proposition 4). Note that NPV emerges in the solution uniquely from axioms on caps, not by assumption. This provides a rigorous support to the conventional view that NPV is the correct evaluation criterion, deduced from reasonable axioms on threshold tests. Moreover, assuming a regularity condition to hold, we show that the set of legal loans is the lower contour set at $r$ of a quasiconvex extension of the IRR to the set of all loans (Proposition 5). This result is consonant with the current state of affairs, according to which the set of legal loans is the lower contour set of the conventional IRR at $r$. We adopt most axioms from Promislow (1997), relaxing the requirement that the set of cash flow streams associated with illegal loans be closed under addition. Note that the dual requirement for the set of legal loans is natural: it guarantees that a lender cannot get around the law and make an illegal loan by decomposing it into several legal ones.

Finally, we suggest a way to resolve ambiguity C by imposing both a ceiling and a floor on lending rates. With this approach, we need not to identify the borrower/lender roles for the parties of a loan contract, by allowing each party speaks for itself. By inspecting whether the rights of each party are respected, we are able to identify the legal status of the contract and, if the status is illegal, determine which party should be brought to justice.

Our findings are general and extend beyond a usury law. As shown in Promislow (1997), Vilensky and Smolyak (1999), Sokolov (2024), and this paper (Proposition 2), the concept of IRR cannot be meaningfully extended to all cash flows; however, this paper demonstrates that its lower and upper contour sets can be (Propositions 3, 4, and 8). This result applies to any setting where one wishes to extend an IRR-based threshold rule to arbitrary cash flows. Applications include lending and deposit rates regulation, investment screening, and capital budgeting, where the standard decision rule accepts a project if its IRR exceeds the hurdle rate.

The rest of the paper is organized as follows. Section 2 contains preliminaries; it introduces the space of cash flow streams we deal with (a loan is identified with the cash flow stream it generates) and describes the structure of net present value functionals on that space. Section 3 presents an axiomatic approach to IRR. It delineates a variety of IRRs, from which one has to be selected to eliminate ambiguity A. Section 4 clarifies ambiguity B by showing how to extend the effective interest rate cap induced by a particular IRR to all loans. In Section 5, implications of the conventional and extended effective interest rate caps are illustrated with examples. Section 6 outlines a few further modifications and generalizations of the concept of effective interest rate cap, as well as the dual concept—effective interest rate floor. It also suggests a way to resolve ambiguity C by considering both a ceiling and a floor on lending rates. All proofs and auxiliary results are provided in the Appendix.

## 2. Preliminaries

We begin with basic definitions and notation. $\mathrm{R}_{++}$, $\mathrm{R}_{--}$, $\mathrm{R}_{+}$, $\mathrm{R}_{-}$, and $\mathrm{R}$ are the sets of positive, negative, nonnegative, nonpositive, and all real numbers, respectively. We use the following notation for one-sided limits and limits at infinity of a function $f:\mathrm{R}\to\mathrm{R}$: $f(t-):=\lim_{\tau\to t-} f(\tau)$, $f(t+):=\lim_{\tau\to t+} f(\tau)$, $f(+\infty):=\lim_{\tau\to+\infty} f(\tau)$, $f(-\infty):=\lim_{\tau\to-\infty} f(\tau)$. By a correspondence $\Gamma$ from a set $\mathrm{X}$ into a set $\mathrm{Y}$, we mean a map from $\mathrm{X}$ into the power set $2^{\mathrm{Y}}$. We write $\Gamma:\mathrm{X}\rightrightarrows\mathrm{Y}$ to distinguish a correspondence from a function from $\mathrm{X}$ to $\mathrm{Y}$.

By a *loan* we mean a function $x:\mathrm{R}_{+}\to\mathrm{R}$ satisfying the following three conditions: (A) $x$ has bounded variation, (B) there is $T\in\mathrm{R}_{+}$ such that $x$ is constant on $[T,+\infty)$, and (C) $x$ is right-continuous. The function $x$ is interpreted as the lender's cumulative (deterministic) cash flow associated with the loan. That is, $x(t)$ is the balance of the lender at time $t$—the difference between cumulative cash inflows and cash outflows to the borrower over the time interval $[0,t]$. We prefer to describe a loan by means of the cumulative (rather than net) cash flow as this setup enables a uniform treatment of discrete- and continuous-time settings. In view of the Jordan decomposition theorem (Monteiro et al. 2018, Theorem 2.1.21, p. 17), condition (A) states that a loan $x$ can be represented in the form $x=x_{+}-x_{-}$, where $x_{+}$ and $x_{-}$ are nondecreasing functions. Such a representation is vital for $x$ to be interpreted as a cumulative cash flow as, by definition, it is the net of cumulative cash inflows and outflows. Condition (B) states that a loan has a finite maturity date. In what follows, the least $T$ satisfying condition (B) is called *the maturity date of* $x$ (by condition (C), the maturity date is well defined). Though real-world loans have discrete cash flows with finitely many transactions, loans with continuous cash flows (e.g., continuously paying annuities) or infinitely many transactions serve as important theoretical constructions. A necessary condition for such construction to be relevant is its ability to approximate a real-world loan. This motivates condition (C): given (A) and (B), condition (C) is necessary and sufficient for $x$ to be an adherent point of the set of discrete loans (in the topological space defined below). By condition (A), $x$ is bounded (Monteiro et al. 2018, Remark 2.1.9, p. 12) and has left limits $x(t-)$, $t\in\mathrm{R}_{++}$ and the limit at infinity $x(+\infty)$ (Monteiro et al. 2018, Corollary 2.1.23, p. 18).

The vector space of all loans, denoted by $\mathrm{L}$, is endowed with the strict locally convex inductive limit topology as follows. Let $\mathrm{L}_{T}$, $T=1,2,\ldots$ be the vector subspace of those loans whose maturity date does not exceed $T$ endowed with the topology of uniform convergence. Topologize $\mathrm{L}$ with the finest locally convex topology such that all canonical injections of $\mathrm{L}_{T}$ into $\mathrm{L}$, $T=1,2,\ldots$ are continuous.[3] We write $0_{\mathrm{L}}$ for the zero vector in $\mathrm{L}$.

[3] See Narici and Beckenstein (2010, Section 12.1) for a review of the strict locally convex inductive limit topology.

For any $\tau \in \mathrm{R}_+$, let $1_\tau$ denote the function on $\mathrm{R}_+$ defined by

$$1_\tau(t) := \begin{cases} 1, t \geq \tau \\ 0, t < \tau \end{cases}.$$

$1_\tau$ is the cash flow representing receiving a monetary unit at time $\tau$. The linear span of $\{1_\tau, \tau \in \mathrm{R}_+\}$, denoted by $\mathrm{D}$, corresponds to the practically relevant case of discrete cash flow streams with finitely many transactions. It can be shown that $\mathrm{D}$ is dense in $\mathrm{L}$ (see part (b) of Lemma 3 in the Appendix). Thus, $\mathrm{L}$ is a natural extension of the practically relevant space of discrete loans.

The topological dual of $\mathrm{L}$ (resp. $\mathrm{L}_T$) is denoted by $\mathrm{L}^*$ (resp. $\mathrm{L}_T^*$). We equip $\mathrm{L}^*$ with the weak$^*$ topology. The polar cone of a set $\mathrm{C} \subseteq \mathrm{L}$ is given by $\mathrm{C}^\circ := \{F \in \mathrm{L}^* : F(x) \leq 0 \ \forall x \in \mathrm{C}\}$. The polar cone of a set $\mathrm{K} \subseteq \mathrm{L}^*$ is defined in a similar fashion, $\mathrm{K}^\circ := \{x \in \mathrm{L} : F(x) \leq 0 \ \forall F \in \mathrm{K}\}$. We let $\mathrm{L}_- := \{x \in \mathrm{L} : x(t) \leq 0 \ \forall t \in \mathrm{R}_+\}$ denote the set of cash flows with the property that the cumulative cash outflow all the time dominates the cumulative cash inflow and write $x \leq y$ if $x - y \in \mathrm{L}_-$.

Following an axiomatic approach to valuation of cash flow streams (Norberg 1990; Promislow 1994; Armerin 2014), by an *NPV functional*, we mean an additive (i.e., $F(x) + F(y) = F(x+y)$ for all $x, y \in \mathrm{L}$) and positive (i.e., $F(\mathrm{L}_-) \subseteq \mathrm{R}_-$) functional $F : \mathrm{L} \to \mathrm{R}$ satisfying the normalization condition $F(1_0) = 1$. We let $\mathcal{NPV}$ denote the set all NPV functionals. Since $\mathrm{L}_-$ has nonempty interior ($-1_0$ is an interior point of $\mathrm{L}_-$), an additive and positive functional on $\mathrm{L}$ is homogeneous and continuous (Jameson 1970, Corollary 3.1.4, p. 81), so $\mathcal{NPV} = \{F \in \mathrm{L}_-^\circ : F(1_0) = 1\}$. Denote by $\mathcal{A}$ the set of all nonincreasing functions $\alpha : \mathrm{R}_+ \to \mathrm{R}_+$ satisfying $\alpha(0) = 1$. It can be shown (see Lemma 4 in the Appendix) that $F \in \mathcal{NPV}$ if and only if there exists $\alpha \in \mathcal{A}$ such that

$$F(x) = x(0) + \int_0^\infty \alpha(t) \mathrm{d}x(t), \tag{1}$$

where the integral is the Kurzweil-Stieltjes integral.[4] For a discrete cash flow stream $x = \sum_{k=0}^{n} x_k 1_{t_k} \in \mathrm{D}$, where $x_k$ is the net cash flow at time $t_k$, Eq. (1) reduces to the familiar discounted sum $F(x) = \sum_{k=0}^{n} \alpha(t_k) x_k$. For a differentiable cumulative cash flow stream $x$ with $x(0) = 0$, Eq. (1) reduces to another familiar expression appeared in continuous-time setup, $F(x) = \int_0^\infty \alpha(t) x'(t) \mathrm{d}t$, where $x'(t)$ is interpreted as the instantaneous net cash flow (the instantaneous rate of payment) at time $t$. As $\alpha(t) = F(1_t)$, i.e., $\alpha(t)$ is the present worth of receiving a monetary unit (the discount factor) at time $t$, elements of $\mathcal{A}$ are hereafter called *discount functions*. We use the notation $F^{(\alpha)}$ for an NPV functional whenever we want to emphasize that it is induced by the discount function $\alpha$ via Eq. (1).

[4] See Monteiro et al. (2018, Chapter 6) for a review of the Kurzweil-Stieltjes integral. For continuous $\alpha$ the integral in Eq. (1) reduces to the Riemann-Stieltjes integral.

For any $x \in \mathrm{L}$ and $r \in \mathrm{R}$, set

$$x_r(\tau) := x(0) + \int_0^{\tau} e^{-rt} \mathrm{d}x(t).$$

The function $x_r : \mathrm{R}_+ \to \mathrm{R}$ represents the cumulative discounted (at the rate $r$) cash flow associated with $x$. Note that $x_r \in \mathrm{L}$ (Monteiro et al. 2018, Corollary 6.5.5, p. 172). Put $F_r(x) := x_r(+\infty)$, $r \in \mathrm{R}$. Since $F_r$ is linear and its restriction to each $\mathrm{L}_T$, $T = 1, 2, \ldots$ is continuous (Monteiro et al. 2018, Theorem 8.2.8, p. 304), we have $F_r \in \mathrm{L}^*$ (Narici and Beckenstein 2010, Theorem 12.2.2, p. 434). Provided that $r \in \mathrm{R}_+$, $F_r$ is the NPV functional induced by the exponential discount function $t \mapsto e^{-rt}$.

## 3. IRR: an axiomatic approach

In this section, we use an axiomatic approach to introduce IRR and describe a maximal (by inclusion) set on which it is well defined. We begin by introducing the following subsets of $\mathrm{L}$:

$$\begin{aligned}
\mathrm{S}_0 &:= \{-c1_t + ce^{\lambda\tau}1_{t+\tau}, (t,\tau,c,\lambda) \in \mathrm{R}_+ \times \mathrm{R}_{++} \times \mathrm{R}_{++} \times \mathrm{R}\}, \\
\mathrm{S}_1 &:= \{\, x \in \mathrm{L} \setminus \{0_{\mathrm{L}}\} : \text{there exists } \lambda \in \mathrm{R} \text{ such that } x_\lambda \in \mathrm{L}_- \text{ and } x_\lambda(+\infty) = 0 \,\}, \\
\mathrm{S}_2 &:= \{\, x \in \mathrm{L} : \text{there exists } \lambda \in \mathrm{R} \text{ such that } \operatorname{sgn} F_r(x) = \operatorname{sgn}(\lambda - r) \text{ for all } r \in \mathrm{R} \,\}, \text{ and} \\
\mathrm{S}_3 &:= \{\, x \in \mathrm{L} \setminus \{0_{\mathrm{L}}\} : \text{there exists } \lambda \in \mathrm{R} \text{ such that } F_r(x)\operatorname{sgn}(\lambda - r) \geq 0 \text{ for all } r \in \mathrm{R} \,\}.
\end{aligned} \tag{2}$$

The set $\mathrm{S}_0$ contains the simplest loans with two transactions—an initial lending and final repayment. $\mathrm{S}_1$ is the set of pure investments introduced in Teichroew et al. (1965, pp. 155–156). The requirement that $x_\lambda$ is nonpositive means that the status of a lender does not change to that of a borrower. $\mathrm{S}_1$ contains as a proper subset the set of conventional investments with only one change of sign in their net cash flow streams. Some authors (Gronchi 1986; Promislow 2015, Section 2.12) argue that IRR is meaningful for pure investments as well as for the dual set, $-\mathrm{S}_1$, called pure borrowings, only. $\mathrm{S}_2$ is the set of loans that have IRR in its most widespread definition. That is, $x \in \mathrm{S}_2$ if $r \mapsto F_r(x)$ has a unique zero, and at this zero, the function changes sign from positive to negative. Finally, the set $\mathrm{S}_3$ contains nonzero loans for which there exists $\lambda \in \mathrm{R}$ such that $r \mapsto F_r(x)$ is nonnegative on $(-\infty, \lambda]$ and nonpositive on $[\lambda, +\infty)$. Note that $\mathrm{S}_0 \subset \mathrm{S}_1 \subset \mathrm{S}_2 \subset \mathrm{S}_3$, where the second inclusion is established in Lemma 5 in the Appendix. Set $\mathrm{D}_k := \mathrm{S}_k \cap \mathrm{D}$, $k = 1, 2, 3$.

Note that for any $x \in \mathrm{S}_k$, $k \in \{0, \ldots, 3\}$, the value $\lambda$ appeared in the definition of $\mathrm{S}_k$ is unique. This is clear for $\mathrm{S}_0$ and $\mathrm{S}_2$. For $\mathrm{S}_1$ this follows from Lemma 5 in the Appendix. For $\mathrm{S}_3$ this comes from the fact that the function $r \mapsto F_r(x)$, $x \neq 0$ is nonzero and real analytic (Widder 1946, Theorem 5a, p. 57), and therefore, it is nonzero on any nonempty open interval (Krantz and Parks 2002, Corollary 1.2.6, p. 14). Let $I_3 : \mathrm{S}_3 \to \mathrm{R}$ be the function that maps each loan $x \in \mathrm{S}_3$ to the value $\lambda$ that appears in the definition of $\mathrm{S}_3$. Denote by $I_k$ (resp. $J_{k+1}$), $k = 0, 1, 2$ the restrictions of $I_3$ to $\mathrm{S}_k$ (resp. $\mathrm{D}_{k+1}$). One can easily verify that $I_k$, $k \in \{0, 1, 2\}$ maps each loan $x \in \mathrm{S}_k$ to the value $\lambda$ that appears in the definition of $\mathrm{S}_k$. That is, $I_2$ is the conventional (logarithmic or continuously compounded) IRR, $I_1$ is the IRR in the sense of Gronchi (1986) and Promislow (2015, Section 2.12), and $I_0$ is the logarithmic rate of return (i.e., $I_0(-a1_t + b1_{t+\tau}) = (1/\tau)\ln(b/a)$).

We define an IRR as a profitability metric whose restriction to $\mathrm{S}_0$ sends each cash flow $-a1_t + b1_{t+\tau} \in \mathrm{S}_0$ to its logarithmic rate of return $(1/\tau)\ln(b/a)$. More formally, we say that a function $E:\mathrm{P}\to\mathrm{R}$, where $\mathrm{S}_0 \subseteq \mathrm{P} \subseteq \mathrm{L}$, is an *IRR on* $\mathrm{P}$ if the following two conditions hold.

*Consistency* (*CONS*): $x \in \mathrm{S}_0 \;\Rightarrow\; E(x) = I_0(x)$.

*Internality* (*INT*): $x, y, x+y \in \mathrm{P} \;\Rightarrow\; \min\{E(x),E(y)\} \le E(x+y) \le \max\{E(x),E(y)\}$.

As usual, we interpret an IRR as a measure of yield. Condition CONS states that an IRR reduces to the logarithmic rate of return for cash flows from $\mathrm{S}_0$. Condition INT relates the yield for a pool of investments with the yields of its components. According to INT, the union of an investment with one of a higher (resp. lower) yield increases (resp. decreases) the yield of the union. In particular, it justifies the following natural guidance: to guarantee the target yield for a pool of investments, it suffices to keep the target for each project in the pool. In capital budgeting, IRR is a standard tool in accept/reject decision making. Namely, an investment is considered weakly profitable and should be accepted if its IRR is greater than or equal to the hurdle rate. Weakly unprofitable investments are defined in a similar manner. Condition INT states that for each given hurdle rate, the union of weakly profitable (resp. unprofitable) investments is weakly profitable (resp. unprofitable). It follows from condition INT that a (lower or upper) semicontinuous IRR on a cone is positively homogeneous of degree zero; that is, it takes no account of the investment size and hence is a relative measure.

We say that an IRR on $\mathrm{P}$ is *strict* if the inequalities in INT are strict whenever $E(x) \ne E(y)$. This stronger version of condition INT, which we refer to as *strict internality* (*S-INT*), was introduced in Vilensky and Smolyak (1999). Condition S-INT is consistent with the conventional definition of IRR $I_2$: the IRR of the union of investments with IRRs of, say, $3\%$ and $7\%$, if it exists, is strictly between $3\%$ and $7\%$.

Our first result shows that an IRR on a sufficiently rich discrete domain, if any, is unique; and a similar result is valid with respect to a continuous IRR on a sufficiently rich "continuous" domain.

**Proposition 1.**

*The following statements hold.*

(a) *Let* $\mathrm{D}_1 \subseteq \mathrm{P} \subseteq \mathrm{D}\setminus\{0_\mathrm{L}\}$. *A function* $E:\mathrm{P}\to\mathrm{R}$ *is an IRR on* $\mathrm{P}$ *if and only if* $\mathrm{P} \subseteq \mathrm{D}_3$ *and* $E$ *is the restriction of* $J_3$ *to* $\mathrm{P}$.

(b) *Let* $\mathrm{S}_1 \subseteq \mathrm{P} \subseteq \mathrm{L}$. *A function* $E:\mathrm{P}\to\mathrm{R}$ *is a continuous IRR on* $\mathrm{P}$ *if and only if* $\mathrm{P} \subseteq \mathrm{S}_3$ *and* $E$ *is the restriction of* $I_3$ *to* $\mathrm{P}$.

Part (a) of Proposition 1 shows that $J_3$ is a unique IRR on $\mathrm{D}_3$, and moreover, it cannot be extended to a larger set, provided that we restrict ourselves to nonzero discrete cash flow streams. In particular, there is no IRR on the set of all cash flows $\mathrm{L}$. Most real-world loans belong to $\mathrm{D}_1$, which justifies the assumption $\mathrm{D}_1 \subseteq \mathrm{P}$ in part (a). Moreover, real-world cash flows are discrete, so part (a) covers the most interesting case. Assuming continuity, we can say more. Part (b) shows that $I_3$ is a unique continuous IRR on $\mathrm{S}_3$, and furthermore, it cannot be extended to a larger set. It follows from the proof that the following result also holds: if $\mathrm{S}_1 \subseteq \mathrm{P} \subseteq \mathrm{L} \setminus \{0_{\mathrm{L}}\}$ and $E:\mathrm{P} \to \mathrm{R}$ is such that the restriction of $E$ to $\mathrm{S}_1$ is $I_1$, then $E$ is an IRR on $\mathrm{P}$ if and only if $\mathrm{P} \subseteq \mathrm{S}_3$ and $E$ is the restriction of $I_3$ to $\mathrm{P}$. The imposed continuity assumption in part (b) is natural: it states that a minor perturbation of a cash flow stream results in a minor change of an IRR. Nevertheless, as noted in Promislow and Spring (1996), it is rather restrictive and implies, e.g., that an IRR on a sufficiently rich domain cannot be a function of the zeros of $r \mapsto F_r(x)$, since they are discontinuous functions of $x$. For instance, the minimal and maximal zeros, the modifications of the conventional IRR advocated, respectively, in Cantor and Lippman (1983) and Bidard (1999), are discontinuous functions of a cash flow stream.

**Remark 1.**

One can consider IRR whose codomain is the extended real line $\overline{\mathrm{R}} := [-\infty, +\infty]$ (rather than $\mathrm{R}$) equipped with the order topology. A function $E:\mathrm{P} \to \overline{\mathrm{R}}$, where $\mathrm{S}_0 \subseteq \mathrm{P} \subseteq \mathrm{L}$, satisfying CONS and INT is called an *extended IRR on* $\mathrm{P}$. For any $x \in \mathrm{L}$, denote by $f_x$ the function on $\mathrm{R}$ defined by $f_x(r) := F_r(x)$. Set $\mathrm{S}_4 := \mathrm{S}_3 \cup \{x \in \mathrm{L} \setminus \{0_{\mathrm{L}}\}: \ f_x$ is either nonnegative or nonpositive$\}$. Let $I_4 : \mathrm{S}_4 \to \overline{\mathrm{R}}$ be the function defined by $I_4(x) := I_3(x)$ if $x \in \mathrm{S}_3$, $I_4(x) := +\infty$ if $f_x$ is nonnegative, and $I_4(x) := -\infty$ if $f_x$ is nonpositive. A minor modification of the proof of Proposition 1 shows that if $\mathrm{D}_1 \subseteq \mathrm{P} \subseteq \mathrm{D} \setminus \{0_{\mathrm{L}}\}$, then a function $E:\mathrm{P} \to \overline{\mathrm{R}}$ is an extended IRR on $\mathrm{P}$ if and only if $\mathrm{P} \subseteq \mathrm{S}_4 \cap \mathrm{D}$ and $E$ is the restriction of $I_4$ to $\mathrm{P}$. Furthermore, if $\mathrm{S}_1 \subseteq \mathrm{P} \subseteq \mathrm{L}$, then a function $E:\mathrm{P} \to \overline{\mathrm{R}}$ is a continuous extended IRR on $\mathrm{P}$ if and only if $\mathrm{P} \subseteq \mathrm{S}_4$ and $E$ is the restriction of $I_4$ to $\mathrm{P}$.

The next proposition provides similar assertions for a strict IRR. Its part (b) with $\mathrm{P} = \mathrm{S}_2$ is essentially due to Vilensky and Smolyak (1999).

**Proposition 2.**

*The following statements hold.*

(a) *Let* $\mathrm{D}_1 \subseteq \mathrm{P} \subseteq \mathrm{D}$. *A function* $E:\mathrm{P} \to \mathrm{R}$ *is a strict IRR on* $\mathrm{P}$ *if and only if* $\mathrm{P} \subseteq \mathrm{D}_2$ *and* $E$ *is the restriction of* $J_2$ *to* $\mathrm{P}$.

(b) *Let* $\mathrm{S}_1 \subseteq \mathrm{P} \subseteq \mathrm{L}$. *A function* $E:\mathrm{P} \to \mathrm{R}$ *is a continuous strict IRR on* $\mathrm{P}$ *if and only if* $\mathrm{P} \subseteq \mathrm{S}_2$ *and* $E$ *is the restriction of* $I_2$ *to* $\mathrm{P}$.

Loosely speaking, Proposition 2 shows that the conventional definition of IRR is the most general one: each strict IRR on a sufficiently rich domain is the restriction of the conventional IRR. It follows from the proof that the following result also holds: if $\mathrm{S}_1 \subseteq \mathrm{P} \subseteq \mathrm{L}$ and $E:\mathrm{P} \to \mathrm{R}$ is such that the restriction of $E$ to $\mathrm{S}_1$ is $I_1$, then $E$ is a strict IRR on $\mathrm{P}$ if and only if $\mathrm{P} \subseteq \mathrm{S}_2$ and $E$ is the restriction of $I_2$ to $\mathrm{P}$.

Since 1950s (Alchian 1955; Solomon 1956), there has been an ongoing debate as to whether an implicit reinvestment assumption is embedded in the IRR method. Specifically, does the IRR technique implicitly assume that the intermediate cash flows generated by an investment project are reinvested at a rate equal to the IRR? Using different logical arguments as well as mathematical or exemplary evidence, the post-1950s academic and practitioner papers mainly reject the reinvestment assumption (Dudley 1972; Keane 1979; Rich and Rose 2014; Magni and Martin, 2025). Proposition 2 presents yet another formal argument against the reinvestment assumption: it shows that the conventional concept of IRR can be deduced from two natural axioms which have nothing to do with the reinvestment (or financing) assumption.

Some textbooks on capital budgeting define IRR simply as a zero of the function $r \mapsto F_r(x)$, $x \in \mathrm{L}$, provided that the zero is unique. Proposition 2 suggests that for an investment (resp. financing) project, this root uniqueness condition should be supplemented at least by the assumption that at the root, the function changes sign from positive to negative (resp. from negative to positive). If $r \mapsto F_r(x)$ has more than one zero, the literature proposes various generalizations of IRR that reduce to the conventional one whenever $r \mapsto F_r(x)$ has one change of sign. For instance, the minimal zero is important as the asymptotic growth rate of a sequence of repeated projects (Cantor and Lippman 1983). In contrast, Bidard (1999) advocated the maximal zero. More involved selection procedures among the zeros were proposed in Hartman and Schafrick (2004) and Weber (2014). A variety of completely different generalizations of IRR were introduced in Promislow and Spring (1996). Propositions 1 and 2 show that these generalizations necessarily violate both versions of the internality condition. The same conclusion holds for the modified IRR (Lin 1976; Beaves 1988; Shull 1992) and the modifications of IRR introduced in Arrow and Levhari (1969) and Bronshtein and Skotnikov (2007) as they reduce to $I_0$ being restricted to $\mathrm{S}_0$. We want to stress that we treat conditions CONS and INT as minimally reasonable for IRR to be relevant for decision making. Put differently, Propositions 1 and 2 are arguments against various generalizations of the conventional definition of IRR, but these results do not assert to use the conventional definition of IRR instead of its restriction to some proper subset. In particular, they do not contradict Herbst (1978) and Gronchi (1986), who provide arguments that IRR is meaningful, respectively, for conventional and pure investments only.

One might prefer to measure yield over $\mathrm{S}_0$ by means of the compound rate of return, $e^{I_0} - 1$, (or by some other order-preserving transformation of $I_0$) rather than by the logarithmic rate of return $I_0$. So a counterpart of the notion of IRR (resp. strict IRR) that reduces to $e^{I_0} - 1$, rather than to $I_0$, on $\mathrm{S}_0$ is of interest. The structure of this counterpart can be derived from Proposition 1 (resp. Proposition 2) and the following observation. Let $\varphi : \mathrm{R} \to \mathrm{R}$ be continuous and strictly monotone. A function $E : \mathrm{P} \to \mathrm{R}$, where $\mathrm{S}_0 \subseteq \mathrm{P} \subseteq \mathrm{L}$, satisfies condition INT (resp. S-INT) and condition CONS with $I_0$ replaced by $\varphi \circ I_0$ if and only if there is an IRR (resp. a strict IRR) $\tilde{E}$ on $\mathrm{P}$ such that $E = \varphi \circ \tilde{E}$.

## 4. An effective interest rate cap

A typical usury law, in its current wording, restricts the effective interest rate and, thus, is only applicable to loans possessing IRR. Given the results of Section 3, in this section, we show how to extend the law to all loans. We follow an axiomatic approach, which is essentially due to Promislow (1997).

A classification of loans into nonusurious (legal) and usurious (illegal) classes can be defined via a correspondence $\mathcal{N}:\mathrm{R}_{+}\rightrightarrows\mathrm{L}$. Given a maximum allowable (logarithmic) effective interest rate $r\in\mathrm{R}_{+}$, $\mathcal{N}(r)$ and $\mathrm{L}\setminus\mathcal{N}(r)$ are interpreted, respectively, as the sets of nonusurious and usurious loans at that rate. The maximum effective interest rate allowable by a usury law is assumed to be nonnegative, so we restrict the domain of $\mathcal{N}$ to $\mathrm{R}_{+}$. We operate with a correspondence from $\mathrm{R}_{+}$ into $\mathrm{L}$ rather than with a single subset of $\mathrm{L}$ since the relevant authorities normally periodically revise the maximum allowable interest rate, and we wish to impose essential restrictions on $\mathcal{N}$. Let $E$ be an IRR on a set $\mathrm{P}$. We say that $\mathcal{N}:\mathrm{R}_{+}\rightrightarrows\mathrm{L}$ is a *cap consistent with* $E$ (an $E$-*cap*, for short) if the following six conditions hold.

(i) For any $x\in\mathrm{P}$, $x\in\mathcal{N}(r)\;\Leftrightarrow\;E(x)\le r$.

(ii) $x\le y\;\&\;y\in\mathcal{N}(r)\;\Rightarrow\;x\in\mathcal{N}(r)$.

(iii) $\mathcal{N}(r)\subseteq\mathcal{N}(s)$ for any $r\le s$.

(iv) $\mathcal{N}(r)+\mathcal{N}(r)\subseteq\mathcal{N}(r)$.

(v) $\mathrm{R}_{++}\mathcal{N}(r)\subseteq\mathcal{N}(r)$.

(vi) $\mathcal{N}(r)$ is closed.

Most of conditions (i)–(vi) are adopted from Promislow (1997). According to condition (i), an $E$-cap is consistent with the current statement of a usury law which labels a loan from $\mathrm{P}$ as usurious or nonusurious, depending on whether its IRR is greater than, or less than or equal to the maximum allowable rate. Condition (ii) states that a loan with a lower lender's cash flow than a nonusurious loan is nonusurious. An equivalent dual condition asserts that $x\le y\;\&\;x\notin\mathcal{N}(r)\;\Rightarrow\;y\notin\mathcal{N}(r)$. That is, a loan with a higher lender's cash flow than a usurious loan is usurious. According to (iii), if the maximum allowable interest rate increases (resp. decreases), then one would expect the loans that were nonusurious (resp. usurious) at the old rate to remain such. By condition (iv), a lender cannot get around the law and make a usurious loan by decomposing it into several nonusurious ones. In contrast to Promislow (1997), we do not require the set $\mathrm{L}\setminus\mathcal{N}(r)$ of usurious loans to be closed under addition, which seems to be a less natural assumption. According to (v), a classification takes no account of the loan size. Note that some countries establish different categories of interest rate caps based on the loan size (as well as the loan term, type of loan, socio-economic characteristics of the borrower, industry, etc.) (Maimbo and Gallegos 2014; Ferrari et al. 2018); condition (v) is debatable in this case. Finally, by (vi), a small perturbation of a usurious loan is usurious. As noted in Section 1, a loan at a usurious interest rate can, by an arbitrary small perturbation, be transformed into a loan that has no IRR in the conventional sense and, therefore, cannot be evaluated by a typical usury law in its current wording. This creates a loophole for unscrupulous lenders to evade the law by making an "almost" usurious loan. Condition (vi) prevents such a situation from arising.

In what follows, we refer to a correspondence $\mathcal{N} : \mathrm{R}_+ \rightrightarrows \mathrm{L}$ simply as a *cap* if it is an $E$-cap for some IRR $E$. Note that a cap is simply another name for an $I_0$-cap. A cap exists: two particular examples of caps are

$$\mathcal{N}_-(r) := \{x \in \mathrm{L} : x_r \in \mathrm{L}_-\} \text{ and } \mathcal{N}_+(r) := \{F_s,\, s \in [r, +\infty)\}^\circ \tag{3}$$

(one can verify that $\mathcal{N}_-$ and $\mathcal{N}_+$ satisfy conditions (i)–(vi) with $E = I_0$; the only nontrivial part—condition (iii) for $\mathcal{N}_-$—follows from part (a) of Lemma 5 in the Appendix). In contrast, given an IRR $E$, an $E$-cap need not exist. For instance, for the IRR $E$ on $\{0_\mathrm{L}\} \cup \mathrm{S}_0$ defined by $E(0_L) = 1$ and $E(x) = I_0(x)$ for all $x \in \mathrm{S}_0$, there is no $E$-cap. Indeed, assume by way of contradiction that a correspondence $\mathcal{N} : \mathrm{R}_+ \rightrightarrows \mathrm{L}$ is an $E$-cap for that $E$. Then, by condition (i), $0_\mathrm{L} \notin \mathcal{N}(0)$, whereas conditions (v) and (vi) imply that $0_\mathrm{L} \in \mathcal{N}(r)$ for all $r \in \mathrm{R}_+$, which is a contradiction.

In Section 3, we justify four IRRs—$J_2$, $I_2$, $J_3$, and $I_3$—on the basis of their uniqueness and nonextendability properties. The next result shows that these IRRs induce the same unique cap.

**Proposition 3.**

*Let $E$ be the restriction of $I_3$ to a set $\mathrm{P}$, where $\mathrm{D}_2 \subseteq \mathrm{P} \subseteq \mathrm{S}_3$. For a correspondence $\mathcal{N} : \mathrm{R}_+ \rightrightarrows \mathrm{L}$, the following conditions are equivalent.*

(a) $\mathcal{N}$ *is an* $E$*-cap.*

(b) $\mathcal{N} = \mathcal{N}_+$, *where* $\mathcal{N}_+$ *is defined in (3).*

(c) $x \in \mathcal{N}(r) \Leftrightarrow$ *for any* $y \in \mathrm{S}_2$ *with* $I_2(y) > r$, *if* $x + y \in \mathrm{S}_2$, *then* $I_2(x+y) \le I_2(y)$.

Proposition 3 shows that there is a unique extension (satisfying several reasonable conditions) of the current statement of a usury law to all loans consistent with the conventional definition of IRR. To some extent, this result is robust to the definition of IRR—$J_2$, $I_2$, $J_3$, or $I_3$. Moreover, it follows from the proof that Proposition 3 remains valid if the set $\mathrm{D}_2$ in the statement is replaced by the set $\{ x \in \mathrm{D}_2 :\ J_2(x)$ is a simple zero of $s \mapsto F_s(x) \}$ (recall that some authors require IRR to be a simple root of the IRR polynomial).[5] Given a maximum allowable interest rate $r$, the cap $\mathcal{N}_+$ labels a loan as usurious if the lender's cash flow has positive NPV at some discount rate $s \ge r$. In particular, if $x(0) < 0$, then $x$ is usurious if and only if the largest zero (if any) of $s \mapsto F_s(x)$ such that at this zero, the function changes sign from positive to negative, exceeds $r$. Thus, for loans whose lender's cash flow starts with an outflow and whose IRR polynomial has simple roots, the cap is consistent (in the sense of condition (i)) with the rule of largest root of the IRR polynomial advocated in Bidard (1999). A less functional but intuitive description of the cap $\mathcal{N}_+$ is given in part (c): given a maximum allowable interest rate $r$, a loan $x$ is usurious if and only if there is a loan $y \in \mathrm{S}_2$ with $I_2(y) > r$ whose union with $x$ increases the IRR. It follows from the proof that Proposition 3 remains valid if the set $\mathrm{S}_2$ and the IRR $I_2$ in part (c) are replaced by $\mathrm{S}_3$ and $I_3$.

According to Proposition 3, given a maximum allowable effective interest rate $r$, legality of a loan $x \in \mathrm{L}$, i.e., $x \in \mathcal{N}_+(r)$, can be verified by checking whether the function $s \mapsto F_s(x)$ is nonpositive on $[r,+\infty)$. If $x \neq 0_{\mathrm{L}}$, then $x \in \mathcal{N}_{+}(r)$ if and only if there is $s_0 \in [r,+\infty)$ such that $F_{s_0}(x) < 0$ and $s \mapsto F_s(x)$ has no zeros of odd multiplicity in $(r,+\infty)$. The answer to the latter question can be reduced to numerical integration (Hungerbühler and Wasem 2018, Corollary 3.4). A simpler procedure is available in the practically relevant case when $x$ lies in the linear span of $\{1_\tau, \tau \in \{0,1,...\}\}$. In this case, $s \mapsto F_s(x)$ can be converted to a polynomial by making the change of variable, $f_x(z) := F_{-\ln z}(x)$. If $x \neq 0_{\mathrm{L}}$ and $f_x$ has no zeros of even multiplicity in the interval $(0, e^{-r})$, then $x \in \mathcal{N}_{+}(r)$ if and only if $x$ starts with an outflow and $f_x$ has no zeros in $(0, e^{-r})$. The latter question in turn can be answered by the use of Sturm's theorem (Basu et al. 2003, Theorem 2.56, p. 48), which gives a procedure for determining the number of distinct real roots of a polynomial in a given interval. This procedure can be turned into a quadratic time algorithm as a function of the degree of the polynomial (Basu et al. 2003, Algorithm 9.2, p. 284). Some care is required if $f_x$ has zeros of even multiplicity in $(0, e^{-r})$. One can handle this case by applying Sturm's procedure iteratively to a polynomial whose zeros are precisely the multiple zeros of $f_x$ (which in turn is a byproduct of Sturm's procedure).

[5] We shall say that a zero $r$ of a differentiable function $f$ is simple if $f'(r) \ne 0$.

Some authors argue that the root uniqueness condition in the form it is used in the definition of $I_2$ is not sufficient to be relevant for decision making. For instance, Gronchi (1986) and Promislow (2015, Section 2.12) assert that IRR is meaningful for pure investments $\mathrm{S}_1$ (as well as for pure borrowings, $-\mathrm{S}_1$) only. Herbst (1978) argues that IRR is a proper measure of return on investment just for conventional investments that have only one change of sign in their net cash flow streams, which is a proper subset of $\mathrm{S}_1$. Our next result characterizes caps consistent with those definitions of IRR. In particular, it describes $I_1$-, $J_1$-, and $I_0$- (i.e., all) caps.

**Proposition 4.**

*Let $E$ be the restriction of $I_1$ to a set $\mathrm{P}$, where $\mathrm{S}_0 \subseteq \mathrm{P} \subseteq \mathrm{S}_1$. For a correspondence $\mathcal{N} : \mathrm{R}_{+} \rightrightarrows \mathrm{L}$, the following conditions are equivalent.*

(a) *$\mathcal{N}$ is an $E$-cap.*

(b) *$\mathcal{N}$ is a cap.*

(c) *There is a correspondence $\mathcal{F} : \mathrm{R}_{+} \rightrightarrows \mathcal{NPV}$ such that for any $r \in \mathrm{R}_{+}$,*

1. *$\mathcal{N}(r) = \mathcal{F}(r)^{\circ}$;*
2. *$F_r \in \mathcal{F}(r)$;*
3. *if $F^{(\alpha)} \in \mathcal{F}(r)$, then $-(\ln \alpha(t))' \geq r$, $t \in \mathrm{R}_{+}$, whenever the left-hand side of the inequality is defined;*
4. *$\mathcal{F}(r) \subseteq \mathcal{F}(s)$ for any $r \geq s$.*

(d) *Conditions (iii)–(vi) hold and for any $r \in \mathrm{R}_{+}$,*

$$\mathcal{N}_{-}(r) \subseteq \mathcal{N}(r) \subseteq \mathcal{N}_{+}(r), \tag{4}$$

*where $\mathcal{N}_{-}$ and $\mathcal{N}_{+}$ are defined in (3).*

All concepts of IRR that appear in the literature reduce to $I_0$ on $\mathrm{S}_0$. Thus, if there is a cap consistent with a particular concept of IRR, then it must be of the form described in Proposition 4. It follows from Proposition 4 that each cap $\mathcal{N}$ is consistent with the current statement of a usury law for loans from $\mathrm{S}_1$: if $x \in \mathrm{S}_1$, then $x \in \mathcal{N}(r)$ if and only if $I_1(x) \le r$. Note that most real-world loans belong to $\mathrm{D}_1$ and all caps classify them identically. Given an IRR $E$ on a set $\mathrm{P}$, this also provides a simple necessary condition for an $E$-cap to exist. Namely, if an $E$-cap exists, then $x \in \mathrm{P} \cap \mathrm{S}_1$ & $I_1(x) > 0 \Rightarrow E(x) = I_1(x)$.

Propositions 3 and 4 show that there is no gap between $I_0$- and $I_1$-caps (in particular, Proposition 4 also describes caps consistent with the IRR on the set of conventional investments that have only one change of sign in their net cash flow streams advocated in Herbst 1978), as well as between $J_2$- and $I_3$-caps, whereas there is a gap between $I_1$- and $I_2$-caps (as well as between $J_1$- and $J_2$-caps). Proposition 4 characterizes a variety of caps. According to part (c), in all of them, the set of nonusurious loans $\mathcal{N}(r)$, $r \in \mathrm{R}_+$ is the polar cone of a collection of NPV functionals whose discount functions meet the requirement that at any date the instantaneous discount rate, if it exists, equals or exceeds $r$ (condition 3). We emphasize that NPV functionals in the solution emerge from the cap axioms rather than by assumption. Given that these axioms contain no explicit assumptions regarding cash flow valuation, this provides a rigorous foundation for the conventional wisdom that NPV is the correct evaluation criterion, deduced from reasonable axioms on threshold tests.

Part (d) of Proposition 4 provides sharp lower and upper bounds on the sets of nonusurious loans; one can treat them, respectively, as necessary and sufficient conditions for a loan to be nonusurious. The upper bound corresponds to the cap $\mathcal{N}_+$ appeared in Proposition 3. The lower bound results in the cap $\mathcal{N}_-$, which can be characterized as follows.

**Lemma 1.**

*For any $r \in \mathrm{R}_+$ and $x \in \mathrm{L}$, the following conditions are equivalent.*

(a) $x \in \mathcal{N}_-(r)$.

(b) $x \in \{F^{(\alpha)} \in \mathcal{NPV}\colon\ -(\ln \alpha(t))' \ge r,\ t \in \mathrm{R}_+$, *whenever the left-hand side of the inequality is defined*$\}^\circ$.

(c) *Either $x = 0_{\mathrm{L}}$ or there is $y \in \mathrm{S}_1$ such that $x \le y$ and $I_1(y) = r$.*

The equivalence of parts (a) and (b) of Lemma 1 easily follows from Proposition 4 if we note that the sets $\mathcal{F}(r) = \{F^{(\alpha)} \in \mathcal{NPV}\colon\ -(\ln \alpha(t))' \ge r$, $t \in \mathrm{R}_+$, whenever the left-hand side of the inequality is defined$\}$, $r \in \mathrm{R}_+$ are the largest (by inclusion) subsets of $\mathcal{NPV}$ satisfying conditions 2–4 in part (c) of Proposition 4. As shown in part (c) of Lemma 1, the cap $\mathcal{N}_-$ is quite intuitive: given a maximum allowable interest rate $r$, a nonzero loan is classified as nonusurious if and only if it is dominated by a pure investment with the IRR $r$.

We say that a cap $\mathcal{N}$ is *stable* if $\mathcal{N}(r) = \{x \in \mathrm{L} : x_r \in \mathcal{N}(0)\}$ $\forall r \in \mathrm{R}_{++}$. To motivate the stability condition, note that most known definitions of IRR, including the conventional one, have the property that if a cash flow $x \in \mathrm{L}$ has the IRR $r$, then $x_s$, $s \in \mathrm{R}$ has the IRR $r - s$. A stable cap requires this type of property to hold for all cash flows: if $x \in \mathcal{N}(r)$ (resp. $x \notin \mathcal{N}(r)$) and $s \le r$, then $x_s \in \mathcal{N}(r-s)$ (resp. $x_s \notin \mathcal{N}(r-s)$). A stable cap is particularly convenient in applications due to its simple structure: it is determined by a single subset of $\mathrm{L}$—$\mathcal{N}(0)$—rather than by a continuum of subsets—$\langle \mathcal{N}(r), r \in \mathrm{R}_+ \rangle$ (note that a cap $\mathcal{N}$ cannot be implemented in practice unless there is a constructive way to define the indexed family $\langle \mathcal{N}(r), r \in \mathrm{R}_+ \rangle$). Examples of stable caps are $\mathcal{N}_-$ and $\mathcal{N}_+$. Our next result describes the general structure of stable caps.

**Corollary 1.**

*For a set* $\mathrm{N} \subseteq \mathrm{L}$, *the following conditions are equivalent.*

(a) *The correspondence* $r \mapsto \{x \in \mathrm{L} : x_r \in \mathrm{N}\}$ *is a stable cap.*

(b) *There is a set* $\Phi \subseteq \mathcal{NPV}$ *such that*

1'. $\mathrm{N} = \Phi^\circ$;

2'. $F_0 \in \Phi$;

3'. *if* $F^{(\alpha)} \in \Phi$, *then so does the NPV functional induced by the discount function* $t \mapsto \alpha(t)e^{-rt}$ *for all* $r \in \mathrm{R}_{++}$.

Corollary 1 shows that a stable cap is determined by a subset $\Phi \subseteq \mathcal{NPV}$ such that $F_0 \in \Phi$ and with every $F^{(\alpha)} \in \Phi$, $\Phi$ also contains the NPV functional with the discount function $t \mapsto \alpha(t)e^{-rt}$ $\forall r \in \mathrm{R}_{++}$. We interpret $\Phi$ as the set of valuation functionals corresponding to a set of feasible economic scenarios. Thus, $x \in \mathrm{N}$ (i.e., $x$ is nonusurious for all $r \in \mathrm{R}_+$) if and only if $x$ is unprofitable (i.e., has nonpositive NPV) in every feasible scenario. For instance, in the case of the cap $\mathcal{N}_-$, we have $\Phi = \mathcal{NPV}$, i.e., all scenarios are feasible. In the case of $\mathcal{N}_+$, we have $\Phi = \{F_r, r \in \mathrm{R}_+\}$ (equivalently, $\Phi$ is the closed convex hull of $\{F_r, r \in \mathrm{R}_+\}$—the set of NPV functionals induced by the set of completely monotone discount functions).

A reasonable requirement on a cap, which is not mentioned among (i)–(vi), is continuity (in some sense); that is, a minor perturbation of a maximum allowable interest rate should result in a minor perturbation of the classification. One natural notion of continuity can be introduced as follows. A cap $\mathcal{N}$ is *continuous* if $\mathcal{N}(r) = \bigcap_{s:s>r} \mathcal{N}(s)$ for all $r \in \mathrm{R}_+$ and $\mathcal{N}(r) = \mathrm{cl}\left( \bigcup_{s:s<r} \mathcal{N}(s) \right)$ for all $r \in \mathrm{R}_{++}$, where $\mathrm{cl}$ is the topological closure operator in $\mathrm{L}$. Intuitively, the first (resp. second) condition in the definition of continuity guarantees that $\mathcal{N}(r)$ does not expand (resp. shrink) dramatically consequent on a small increase (resp. decrease) in $r$. In general, a cap need not be continuous. However, as shown in the next lemma, stable caps are continuous. In particular, so are $\mathcal{N}_-$ and $\mathcal{N}_+$.

**Lemma 2.**

*A stable cap is continuous.*

According to the current generic formulation of a usury law, the set of nonusurious loans is the corresponding lower contour set of the conventional IRR. Our next result establishes a similar connection between a continuous cap and the collection of lower contour sets of a functional on $\mathrm{L}$. Namely, given an IRR $E$, a continuous $E$-cap maps a maximum allowable interest rate to the corresponding lower contour set of an extension of the positive part of $E$ to the set of all loans $\mathrm{L}$.

In order to formulate this result, we introduce the following definition. Given an IRR $E : \mathrm{P} \to \mathrm{R}$, a function $\overline{E} : \mathrm{L} \to \mathrm{R}_+ \cup \{+\infty\}$ is a *refinement of* $E$ if the following five conditions hold.

(I) $\overline{E}(x) = \max\{0, E(x)\}$, whenever $x \in \mathrm{P}$.

(II) $x \le y \;\Rightarrow\; \overline{E}(x) \le \overline{E}(y)$.

(III) $\overline{E}(\lambda x) = \overline{E}(x)$ for all $x \in \mathrm{L}$ and $\lambda \in \mathrm{R}_{++}$.

(IV) For every $r \in \mathrm{R}_+$, the set $\{x \in \mathrm{L} : \overline{E}(x) \le r\}$ is closed and convex.

(V) For every $r \in \mathrm{R}_{++}$, $\mathrm{cl}(\{x \in \mathrm{L} : \overline{E}(x) < r\}) = \{x \in \mathrm{L} : \overline{E}(x) \le r\}$.

By construction, $\overline{E}$ is an extension of the positive part of $E$ (condition (I)). It is nondecreasing (condition (II)), positively homogeneous of degree zero (condition (III)), lower semicontinuous, and quasiconvex (condition (IV)). Finally, $\overline{E}$ satisfies a version of local nonsatiation (condition (V)), which rules out "thick" level sets. We interpret a refinement of $E$ as a lower semicontinuous extension of the positive part of $E$ that preserves the second (but not necessarily the first) inequality in condition INT. Our next result describes the general structure of continuous caps.

**Proposition 5.**

*Let $E$ be an IRR. For a correspondence $\mathcal{N} : \mathrm{R}_+ \rightrightarrows \mathrm{L}$, the following conditions are equivalent.*

(a) *$\mathcal{N}$ is a continuous $E$-cap.*

(b) *There is a refinement $\overline{E}$ of $E$ such that for all $r \in \mathrm{R}_+$, $\mathcal{N}(r) = \{x \in \mathrm{L} : \overline{E}(x) \le r\}$.*

Given an IRR $E$, Proposition 5 shows that a continuous $E$-cap maps $\mathrm{R}_+$ onto the collection of lower contour sets of a refinement of $E$. Moreover, it follows from the proof that the map that sends a continuous $E$-cap $\mathcal{N}$ to the function from $\mathrm{L}$ to $\mathrm{R}_+ \cup \{+\infty\}$ given by $x \mapsto \inf\{r \in \mathrm{R}_+ : x \in \mathcal{N}(r)\}$ (with the convention $\inf \varnothing = +\infty$) defines a bijection from the set of continuous $E$-caps onto the set of refinements of $E$; the inverse map sends a refinement $\overline{E}$ of $E$ to the correspondence $r \mapsto \{x \in \mathrm{L} : \overline{E}(x) \le r\}$. For instance, a stable cap $r \mapsto \{x \in \mathrm{L} : x_r \in \mathrm{N}\}$, $\mathrm{N} \subseteq \mathrm{L}$ is continuous (Lemma 2); thus, by Proposition 5, its image is the collection of the lower contour sets of the refinement of $I_0$ defined by $x \mapsto \inf\{r \in \mathrm{R}_+ : x_r \in \mathrm{N}\}$. By way of illustration, consider the stable cap $\mathcal{N}_-$. Its image is the collection of the lower contour sets of the refinement of $I_0$ defined by $W(x) := \inf\{r \in \mathrm{R}_+ : x_r \in \mathrm{L}_-\}$. In view of Lemma 1, $W$ can also be represented as $W(0_\mathrm{L}) = 0$ and $W(x) = \inf\{I_1(y) : y \in \mathrm{S}_1, x \le y, I_1(y) \ge 0\}$, $x \ne 0_\mathrm{L}$. That is, $W$ is the modification of IRR introduced in Bronshtein and Skotnikov (2007), which maps each nonzero cash flow $x$ to the least IRR over the set of pure investments dominating $x$. As noted in Bronshtein (2007), one can prove that $W$ coincides on $\{x \in \mathrm{L} : W(x) \in \mathrm{R}_{++}\}$ with the modification of IRR introduced in Arrow and Levhari (1969). As another illustration, combining Propositions 3 and 5, we obtain that the conventional IRR $I_2$ has a unique refinement, $x \mapsto \inf\{r \in \mathrm{R}_+ : F_s(x) \le 0 \;\forall s \in [r, +\infty)\}$. For a loan that starts with an outflow, this refinement reduces to the largest root of the IRR polynomial such that at this root the polynomial changes sign from positive to negative, i.e., to the modification of IRR advocated in Bidard (1999).

## 5. Implications

The examples below illustrate the application of caps.

1°. Recall that a typical usury law, in its current wording, is unable to evaluate a loan with an application fee (as well as any other lender fee charged before a loan is processed) as the associated cash flow stream has no IRR. It follows from Proposition 4 that every cap prohibits application fees. Indeed, if a lender's cash flow $x$ starts with an inflow, then $F_s(x) > 0$ for sufficiently large $s$, so $x$ is classified as usurious for any maximum allowable interest rate. In contrast, every cap classifies a loan $x$, whose lender's cash flow starts with an outflow, as nonusurious for sufficiently large maximum allowable interest rate as $x_r \in \mathrm{L}_-$ for $r$ large enough. Hence, one possible way to make an application fee legal is to charge it at the loan issue date, which is already standard practice in some countries.

2°. Banks in Russia offer a loan with an option that the bank reduces the interest rate, say, from 7% to 4% and refunds the difference after the loan is fully repaid along with the interest, provided that the borrower makes all loan repayments on time, according to the loan repayment schedule, and meets some other requirements.[6] Let $x$ ($y$) be the lender's cumulative cash flow stream associated with the loan with (without) refund. Provided that the ceiling is, say, 10%, the usury law, in its current wording, authorizes $y$, but is unable to evaluate $x$: as $x$ ends with an outflow, we have $F_s(x) < 0$ for sufficiently small $s$, so $x$ has no IRR. In contrast, given a cap, if $y$ is nonusurious, then so is $x$ by condition (ii). This result is intuitive as $x$ is more favorable to the borrower than $y$. A similar conclusion holds for any loan accompanied by a refund. As noted in Section 1, a refund occurs with regular frequency in particular types of loans or may be caused by force majeure. For instance, in September 2022, the U.S. Department of Education announced that borrowers who held U.S. federal student loans and kept making payments during the COVID-19 pandemic, were eligible for a refund.[7] Though later this initiative of President Joe Biden's administration was blocked, it would potentially affect more than 9 million borrowers.

3°. Consider a line of credit consisting of loans $x$ and $y$. Following the spirit of a usury law, if the law authorizes $x$ and $y$, then it also has to authorize the whole line of credit $x + y$ as the lender can make it by decomposing into $x$ and $y$. However, in general, this is not the case for the current wording of a usury law. Indeed, one can easily construct loans $x, y \in \mathrm{S}_2$ (or even $x, y \in \mathrm{S}_0$) such that $x + y \notin \mathrm{S}_2$, i.e., $x$ and $y$ has the IRR in the conventional sense, whereas the IRR for $x + y$ is undefined. Thus, for sufficiently large maximum allowable interest rate, the usury law, in its current wording, authorizes $x$ and $y$, but is unable to evaluate $x + y$. In contrast, by condition (iv), given a cap, if $x$ and $y$ are nonusurious, then so is $x+y$. By way of illustration, consider 1-year loans $x=-1_0+5\cdot 1_1$ and $y=-1000\cdot 1_6+1500\cdot 1_7$. Then $x+y$ is a line of credit with (dramatically) variable interest rate, from 400% in the first period to 50% in the seventh period. Loans $x$ and $y$ have the IRRs, $I_0(x)=\ln(5/1)\approx 1.61$ and $I_0(y)=\ln(1500/1000)\approx 0.41$, whereas $x+y$ has not: the IRR polynomial for $x+y$, i.e., $s\mapsto F_s(x+y)$, has three roots, 0.44, 1.11, and 1.55. So the usury law is inapplicable to $x+y$. In contrast, if a cap authorizes $x$ and $y$, then it also authorizes $x+y$ (condition (iv)).

[6] Pochta Bank's "Guaranteed Rate" option (in Russian): https://www.pochtabank.ru/news/709062, accessed on June 26, 2025. Renaissance Bank's "Want 0" option (in Russian): https://rencredit.ru/single/furry-zero/, accessed on June 26, 2025. Promsvyazbank's "0 is better" option (in Russian): https://ib.psbank.ru/store/products/consumer-loan, accessed on June 26, 2025.

[7] Federal Student Loan Debt Relief: https://studentaid.gov/debt-relief-announcement/one-time-cancellation, accessed on June 26, 2025.

4°. As noted in Section 1, a loan at a usurious interest rate can, by an arbitrary small perturbation, be transformed into a loan that has no unique IRR and, therefore, cannot be evaluated by a typical usury law in its current wording. This creates a loophole for unscrupulous lenders to evade the law. In contrast, a cap requires the set of usurious loans to be open (condition (vi)), and therefore, it has no such loophole. For instance, turning back to the example from Canada's anti-usury legislation in Section 1, consider loans $x=-100\cdot 1_0+170\cdot 1_1$ and $z_c=-100\cdot 1_0+170\cdot 1_1-c\cdot 1_{366/365}$, $c>0$. Since the Criminal code of Canada prohibits lending money at an effective rate exceeding 60%, it treats $x$ as usurious but is unable to evaluate $z_c$ as it has no IRR. At least for small $c>0$, this is counterintuitive as $z_c$ is a minor perturbation of the usurious loan $x$. In contrast, given the maximum allowable logarithmic interest rate $r=\ln(1.6)$, every cap prohibits $x$ (as $F_r(x)>0$) and $z_c$, provided that $c<10.013$ (as $F_r(z_c)>0$, whenever $c<10.013$). Moreover, there are caps (e.g., $\mathcal{N}_-$) that prohibits $z_c$ for all $c$.

## 6. Modifications and extensions

In this section, we outline a few modifications and extensions of the concept of $E$-cap introduced in Section 4.

1°. *Weak effective interest rate caps*. Condition (iii) in the definition of an $E$-cap appears to be the most open to criticism. It becomes particularly debatable if a maximum allowable interest rate is not assumed to vary as, e.g., in the case of Islamic banking (Sharia prohibits riba—generally defined as interest paid on loans—which formally results in the fixed zero maximum allowable interest rate). In this subsection, we outline the counterparts of Propositions 3 and 4 that correspond to the omission of condition (iii).

Given an IRR $E:\mathrm{P}\to\mathrm{R}$, we say that a correspondence $\mathcal{N}:\mathrm{R}_+\rightrightarrows\mathrm{L}$ is a *weak* $E$*-cap* if it satisfies conditions (i), (ii), (iv)–(vi). A correspondence $\mathcal{N}:\mathrm{R}_+\rightrightarrows\mathrm{L}$ is called a *weak cap* if it is a weak $E$-cap for some IRR $E$. The next two results are counterparts of Propositions 3 and 4 for weak caps.

**Proposition 6.**

*Let* $E$ *be the restriction of* $I_2$ *to a set* $\mathrm{P}$*, where* $\mathrm{D}_2\subseteq\mathrm{P}\subseteq\mathrm{S}_2$*. For a correspondence* $\mathcal{N}:\mathrm{R}_+\rightrightarrows\mathrm{L}$*, the following conditions are equivalent.*

(a) $\mathcal{N}$ *is a weak* $E$*-cap.*

(b) *For any* $r\in\mathrm{R}_+$*,* $\mathcal{N}(r)$ *is a closed convex cone satisfying*

$$\mathcal{N}_+(r) \subseteq \mathcal{N}(r) \subseteq \{F_r\}^\circ. \tag{5}$$

**Proposition 7.**

*Let* $E$ *be the restriction of* $I_1$ *to a set* $\mathrm{P}$*, where* $\mathrm{S}_0 \subseteq \mathrm{P} \subseteq \mathrm{S}_1$*. For a correspondence* $\mathcal{N} : \mathrm{R}_+ \rightrightarrows \mathrm{L}$*, the following conditions are equivalent.*

(a) $\mathcal{N}$ *is a weak* $E$*-cap.*

(b) $\mathcal{N}$ *is a weak cap.*

(c) *There is a correspondence* $\mathcal{F} : \mathrm{R}_+ \rightrightarrows \mathcal{NPV}$ *such that for any* $r \in \mathrm{R}_+$*,*

1. $\mathcal{N}(r) = \mathcal{F}(r)^\circ$*;*
2. $F_r \in \mathcal{F}(r)$*;*
3. *if* $F^{(\alpha)} \in \mathcal{F}(r)$*, then* $-(\ln \alpha(t))' \geq r$*,* $t \in \mathrm{R}_+$*, whenever the left-hand side of the inequality is defined.*

(d) *For any* $r \in \mathrm{R}_+$*,* $\mathcal{N}(r)$ *is a closed convex cone satisfying*

$$\mathcal{N}_-(r) \subseteq \mathcal{N}(r) \subseteq \{F_r\}^\circ. \tag{6}$$

Given a maximum allowable interest rate $r$ and a loan $x$, it follows from Proposition 7 that for a weak cap, $F_r(x) \leq 0$ (resp. $x_r \in \mathrm{L}_-$) is an easily verified necessary (resp. sufficient) condition for $x$ to be nonusurious. In contrast to a cap, a weak cap does not necessarily make illegal a lender fee charged before a loan is processed. An example of a weak cap (which is not a cap) is $r \mapsto \{F_r\}^\circ$. Under this cap, given a maximum allowable interest rate $r$, a loan $x \in \mathrm{L}$ is classified as nonusurious if and only if $F_r(x) \leq 0$. For instance, if $r = 0$ (the case of Islamic banking), then $x$ is nonusurious if and only if $x(+\infty) \leq 0$. The correspondence $r \mapsto \{F_r\}^\circ$ is also the only weak cap for which the sets of usurious loans are convex cones (Sokolov 2024, Example 1), that is, the union of usurious loans is usurious.

2°. *Restricted effective interest rate caps*. The relevant authorities often choose the maximum allowable interest rate from some feasible set $\mathrm{C} \subseteq \mathrm{R}_+$. For instance, $\mathrm{C}$ can be a singleton as in the case of Islamic banking, where $\mathrm{C} = \{0\}$; $\mathrm{C}$ can be discrete as interest rates are usually measured in basis points; $\mathrm{C}$ can be bounded away from zero or bounded from above by a reasonable threshold. In this case, one can modify the exposition by defining a cap as a correspondence from $\mathrm{C}$ into $\mathrm{L}$ satisfying conditions (i)–(vi). This modification, which we call an r-cap, is somewhere between the notions of cap and weak cap in the sense that the restriction of a cap to $\mathrm{C}$ is an r-cap and each r-cap can extended to a weak cap. For instance, if $\mathrm{C}$ is bounded from above, then an r-cap does not necessarily make illegal a lender fee charged before a loan is processed. With minor adjustments to the proofs, one can obtain the analogues of Propositions 3 and 4 for r-caps. In particular, Proposition 4 continues to hold with obvious modifications and $\mathcal{N}_+(r)$ in (4) replaced by $\{F_s,\ s \in [r, +\infty) \cap \mathrm{C}\}^\circ$. We do not elaborate on this.

3°. *Relative effective interest rate caps*. Some countries use relative interest rate ceilings defined as a certain spread over a benchmark rate, which is typically either the central bank's policy rate or an average market interest rate (Ferrari et al. 2018; Calice et al. 2020). The notion of $E$-cap can easily be modified to implement this type of interest rate control as follows. By a *benchmark rate* we mean a locally (Lebesgue) integrable function $b:\mathrm{R}_+ \to \mathrm{R}$, where $b(t)$ is interpreted as the realized instantaneous benchmark rate at time $t$. Given a benchmark rate $b$, to each $x \in \mathrm{L}$ we assign the loan $x^{(b)} \in \mathrm{L}$ by

$$x^{(b)}(\tau) := x(0) + \int_0^{\tau} \exp\left( -\int_0^{t} b(s)\mathrm{d}s \right) \mathrm{d}x(t) . \quad (7)$$

[8]

By construction, $x^{(b)}$ is the cumulative discounted—at the benchmark rate—cash flow associated with $x$. Given a benchmark rate $b$ and an IRR $E$, a correspondence $\mathcal{N}^{(b)}:\mathrm{R}_+ \rightrightarrows \mathrm{L}$ is a *relative* $E$*-cap* if there is an $E$-cap $\mathcal{N}$ such that $\mathcal{N}^{(b)}(r) = \{x : x^{(b)} \in \mathcal{N}(r)\}$ for all $r \in \mathrm{R}_+$. We interpret $\mathcal{N}^{(b)}(r)$ as the set of loans whose excess return earned by the lender over the benchmark rate does not exceed $r$. Put differently, a relative $E$-cap is an $E$-cap applied to the excess return earned over the benchmark rate. For instance, in the case of a constant benchmark rate, $b(t) = \lambda \geq 0 \;\; \forall t \in \mathrm{R}_+$, and a stable $E$-cap $\mathcal{N}$, we have $x^{(b)} = x_\lambda$ and $\mathcal{N}^{(b)}(r) = \{x : x_\lambda \in \mathcal{N}(r)\} = \mathcal{N}(\lambda + r) \;\; \forall r \in \mathrm{R}_+$.

4°. *Effective interest rate floors*. A floor on lending rates is a dual type of interest rate control that enables banks to maintain a minimum margin level. The generic formulation of this interest rate control tool suffers from the ambiguities, similar to A and B. The idea of Section 4 can be applied, with obvious modifications, to resolve these ambiguities and correctly define a floor on lending rates. Namely, given an IRR $E:\mathrm{P} \to \mathrm{R}$, we say that a correspondence $\mathcal{N}:\mathrm{R}_+ \rightrightarrows \mathrm{L}$ is an $E$*-floor* if it satisfies conditions (iv)–(vi) in the definition of an $E$-cap and the following three conditions.

(i)' For any $x \in \mathrm{P}$, $x \in \mathcal{N}(r) \;\Leftrightarrow\; E(x) \geq r$.

(ii)' $x \leq y \;\&\; x \in \mathcal{N}(r) \;\Rightarrow\; y \in \mathcal{N}(r)$.

(iii)' $\mathcal{N}(r) \subseteq \mathcal{N}(s)$ for any $r \geq s$.

Given a minimum allowable (logarithmic) effective interest rate $r \in \mathrm{R}_+$, $\mathcal{N}(r)$ and $\mathrm{L} \setminus \mathcal{N}(r)$ are interpreted, respectively, as the sets of legal and illegal loans at that rate. The interpretation of conditions (i)'–(iii)' and (iv)–(vi) is similar to that of conditions (i)–(vi) in the definition of an $E$-cap and is omitted. One can study $E$-floors in the manner of Section 4. Here we only present the counterpart of the key result of Proposition 4.

**Proposition 8.**

*Let* $E$ *be the restriction of* $I_1$ *to a set* $\mathrm{P}$, *where* $\mathrm{S}_0 \subseteq \mathrm{P} \subseteq \mathrm{S}_1$. *For a correspondence* $\mathcal{N}:\mathrm{R}_+ \rightrightarrows \mathrm{L}$, *the following conditions are equivalent.*

(a) $\mathcal{N}$ *is an* $E$*-floor.*

(b) *There is a correspondence* $\mathcal{F}:\mathrm{R}_+ \rightrightarrows \mathcal{NPV}$ *such that for any* $r \in \mathrm{R}_+$,

1. $\mathcal{N}(r) = -\mathcal{F}(r)^{\circ} = \{x \in \mathrm{L} : F(x) \geq 0 \;\forall F \in \mathcal{F}(r)\}$;

---

[8] In practice, if the instantaneous logarithmic rate of return of a benchmark serves as a benchmark rate, the realized instantaneous benchmark rate may be unobserved. In this case, one can replace $\exp\left( -\int_0^{t} b(s)\mathrm{d}s \right)$ in (7) with $B(0)/B(t)$, where $B(t)$ is the value of the benchmark at time $t$.

2. $F_r \in \mathcal{F}(r)$;

3. *if* $F^{(\alpha)} \in \mathcal{F}(r)$, *then the function* $t \mapsto e^{rt}\alpha(t)$ *is nondecreasing;*

4. $\mathcal{F}(r) \subseteq \mathcal{F}(s)$ *for any* $r \le s$.

For every IRR $E$, an $E$-floor is also an $I_0$-floor; so Proposition 8 describes of all floors. Given $\alpha \in \mathcal{A}$ and $r \in \mathrm{R}_+$, it can be shown that the function $t \mapsto e^{rt}\alpha(t)$ is nondecreasing if and only if $\alpha$ is positive, $\ln\alpha$ is absolutely continuous, and $-(\ln\alpha(t))' \le r$, whenever the derivative in the left-hand side of the inequality exists (see Lemma 7 in the Appendix). That is, for every IRR $E$ and an $E$-floor $\mathcal{N}$, the set of legal loans $\mathcal{N}(r)$, $r \in \mathrm{R}_+$ is the dual cone of a collection of NPV functionals whose instantaneous discount rates do not exceed $r$. If $E$ is the restriction of $I_3$ to a set $\mathrm{P}$, where $\mathrm{S}_0 \subseteq \mathrm{P} \subseteq \mathrm{S}_3$, then an example of an $E$-floor is given by $r \mapsto \{x \in \mathrm{L} : F_s(x) \ge 0 \ \forall s \in [0, r]\}$.

A floor on deposit rates—a frequent dual type of interest rate control around the world (Calice et al. 2020)—suffers from the same ambiguities. The result of this subsection can be applied without any changes to resolve these ambiguities, provided that an element of $\mathrm{L}$ is interpreted as a cash flow stream of the owner of the deposit account.

5°. *Identifying borrower versus lender when roles blur.* When loan advances and repayments alternate in time, the respective roles of the borrower and lender can be blurred. This brings us to the issue of which of the parties of a loan contract a usury law is addressed to (ambiguity C from Section 1). To illustrate the problem, assume that Y and Z sign a contract, according to which Y receives from Z the transaction $y = 1_0 - 2 \cdot 1_1 + 1_2$. One might query who is the lender and who is the borrower in this contract. If Y is treated as the "lender" (say, because the transaction for Y ends with an inflow), then for every maximum allowable interest rate, $\mathcal{N}_+$ classifies the transaction as usurious since $F_s(y) > 0$ for sufficiently large $s$. In contrast, if Z is treated as the "lender" (say, because the transaction for Z starts with an outflow), then for every maximum allowable interest rate, $\mathcal{N}_+$ labels the transaction as nonusurious since $-y \in \mathcal{N}_+(0)$. Thus, the parties of the contract can potentially manipulate the roles of the borrower and lender to evade the law.

One possible solution to this issue is to protect both sides of a loan contract (rather than only the borrower) from usury. This can be implemented by imposing both a floor and a ceiling on lending rates (note that even if a policymaker prefers not to regulate the minimum level of margin, there is a natural floor on lending rates—0%). Given an IRR $E$, an $E$-floor $\mathcal{N}$, an $E$-cap $\mathcal{N}'$, and the minimum and maximum allowable (logarithmic) interest rates—$r$ and $r'$—satisfying $0 \le r \le r'$, a transaction $x \in \mathrm{L}$ is classified as *nonusurious* if either $x$ or $-x$ belongs to $\mathcal{N}(r) \cap \mathcal{N}'(r')$ and *usurious* otherwise. By definition, the manipulation of the roles of the borrower and lender does not affect the result of this classification. Moreover, the classification is capable to identify the actual roles, provided that the transaction is nonzero and nonusurious (this stems from the fact that if $x \ne 0_{\mathrm{L}}$ and $x \in \mathcal{N}(r) \cap \mathcal{N}'(r')$, then $-x \notin \mathcal{N}(r) \cap \mathcal{N}'(r')$). If $x$ is usurious and $x \in \mathcal{N}(r) \cap (\mathrm{L} \setminus \mathcal{N}'(r'))$ or $-x \in (\mathrm{L} \setminus \mathcal{N}(r)) \cap \mathcal{N}'(r')$, then the party of the contract that has $x$ as the cash flow should be brought to justice as the rights of this party are respected, whereas the rights of the other party are violated. Namely, if $x \in \mathcal{N}(r) \cap (\mathrm{L} \setminus \mathcal{N}'(r'))$ (resp.

$-x \in (\mathrm{L} \setminus \mathcal{N}(r)) \cap \mathcal{N}'(r')$), then we treat the party of the contract having $x$ as the lender (resp. borrower) whose rights are respected, whereas the rights of the other side are violated. It may happen that neither of the parties of a usurious transaction is brought to justice; this means that the rights of the both parties are violated.

By way of illustration, consider the IRR $I_2$, the $I_2$-floor $\mathcal{N}(r) = \{x \in \mathrm{L} : F_s(x) \geq 0 \ \forall s \in [0,r]\}$ $\forall r \in \mathrm{R}_+$, the $I_2$-cap $\mathcal{N}' = \mathcal{N}_+$, and the minimum and maximum allowable interest rates of 0% ($r = 0$) and 60% ($r' = \ln(1.6)$), respectively. Then the transaction $y = 1_0 - 2 \cdot 1_1 + 1_2$ (which Y receives from Z in the example above) is classified as nonusurious since $-y \in \mathcal{N}(r) \cap \mathcal{N}'(r')$. Moreover, Y is identified as the borrower and Z as the lender. Now assume that the minimum and maximum allowable interest rates are 3% ($r = \ln(1.03)$) and 60% ($r' = \ln(1.6)$), respectively. Then the transaction $y$ is classified as usurious since neither $y$ nor $-y$ belongs to $\mathcal{N}(r) \cap \mathcal{N}'(r')$. As $y \in \mathcal{N}(r) \cap (\mathrm{L} \setminus \mathcal{N}'(r'))$ and $-y \in (\mathrm{L} \setminus \mathcal{N}(r)) \cap \mathcal{N}'(r')$, whereas $y \notin (\mathrm{L} \setminus \mathcal{N}(r)) \cap \mathcal{N}'(r')$ and $-y \notin \mathcal{N}(r) \cap (\mathrm{L} \setminus \mathcal{N}'(r'))$, the rights of Y are respected and the rights of Z are violated; therefore, Y should be brought to justice.

## 7. Conclusion

A usury law is vague for loans whose cash flow streams have no IRR. In this paper, we use an axiomatic approach to extend the statement of a usury law to all loans. We show that there is a unique extension consistent with the conventional definition of IRR (Proposition 3). Our findings suggest to modify the wording of a usury law as follows: given a maximum allowable interest rate $r$, a loan is usurious if and only if the lender's cash flow has positive NPV at some discount rate $s \geq r$. This modification does not explicitly refer to a particular concept of IRR and therefore eliminates ambiguities A and B of the current generic wording of a typical usury law noted in Section 1. This modification is rather restrictive. For instance, it makes illegal an application fee as well as any other lender fee charged before a loan is processed. A possible way to make an application fee legal is to charge it at the loan issue date. We also derive usury laws consistent with more conservative definitions of IRR (Proposition 4). Specifically, we reveal the generic geometric structure of the set of legal loans: given a maximum allowable interest rate $r$, irrelevantly of the definition of IRR chosen, the set of legal loans is the polar cone of NPV functionals with instantaneous discount rates at least $r$. Our results are general, and apply to any setting where one wishes to extend an IRR-based threshold rule to arbitrary cash flows.

A usury law is also ambiguous whenever the roles of the borrower and lender in a loan contract are blurred (ambiguity C). We suggest a way to resolve this ambiguity by considering both a ceiling and a floor on lending rates.

Our findings also clarify the concept of IRR. We axiomatize the conventional definition of IRR by showing that it is the only extension of the rate of return (defined over the set of investment operations with two transactions, an initial outlay and a final inflow) satisfying a natural axiom. Moreover, the conventional definition of IRR cannot be extended to a larger domain (Proposition 2).

The investment appraisal literature offers a variety of profitability metrics in addition to IRR—the profitability index and the (discounted) payback period, to name a few. Since most of those metrics are partial in the sense that there are cash flow streams for which the metric is undefined,[9] the formulation of the corresponding threshold rule suffers from the drawback similar to ambiguity B in Section 1. As in Sections 4 and 6, we can define caps and floors consistent (in the sense of conditions (i) and (i)') with those profitability metrics.[10] It is an interesting direction for future research to characterize those caps and floors.

## 8. Appendix: auxiliary results and proofs

Throughout this section we use the following notation. For a function $f:\mathrm{R}_+ \to \mathrm{R}$, denote by $\|f\| := \sup_{t\in \mathrm{R}_+} |f(t)|$ its supremum norm and by $\|f\|_a^b$ and $V_a^b(f)$, respectively, the supremum norm and the total variation of $f$ on the interval $[a,b]$.

**Lemma 3.**

*The following statements hold.*

(a) *For any neighborhood* $\mathrm{U}$ *of the origin in* $\mathrm{L}$ *and a natural number* $T$, *there is* $\varepsilon > 0$ *such that* $\{x \in \mathrm{L}_T : \|x\| < \varepsilon\} \subset \mathrm{U}$.

(b) $\mathrm{D} \cap \mathrm{L}_T$, $T = 1,2,\ldots$ *is dense in* $\mathrm{L}_T$; $\mathrm{D}$ *is dense in* $\mathrm{L}$.

(c) *For any* $r \in \mathrm{R}$, *the map* $x \mapsto x_r$ *is a linear self-homeomorphism of* $\mathrm{L}$.

(d) *For any* $r \in \mathrm{R}$, *the map that sends each* $F \in \mathrm{L}^*$ *to the functional* $x \mapsto F(x_r)$ *is a linear self-homeomorphism of* $\mathrm{L}^*$.

(e) *For any* $x \in \mathrm{L}$, *the map from* $\mathrm{R}$ *to* $\mathrm{L}$ *given by* $r \mapsto x_r$ *is continuous.*

**Proof.**

(a). By definition of the strict locally convex inductive limit topology, there exists a convex, balanced, and absorbing neighborhood $\mathrm{V}$ of the origin in $\mathrm{L}$ such that $\mathrm{V} \subseteq \mathrm{U}$ and $\mathrm{V}_T := \mathrm{V} \cap \mathrm{L}_T$ is a neighborhood of the origin in $\mathrm{L}_T$. Therefore, there is $\varepsilon > 0$ such that $\{x \in \mathrm{L}_T : \|x\| < \varepsilon\} \subset \mathrm{V}_T \subset \mathrm{V} \subseteq \mathrm{U}$.

(b). $\mathrm{D} \cap \mathrm{L}_T$ is dense in $\mathrm{L}_T$ (Monteiro et al. 2018, p. 82). To show that $\mathrm{D}$ is dense in $\mathrm{L}$, pick a nonempty open subset $\mathrm{V}$ of $\mathrm{L}$. Since $\mathrm{L} = \bigcup_T \mathrm{L}_T$, there is $T$ such that $\mathrm{V} \cap \mathrm{L}_T \neq \varnothing$. As $\mathrm{D} \cap \mathrm{L}_T$ is dense in $\mathrm{L}_T$, for this $T$, the open (in $\mathrm{L}_T$) set $\mathrm{V} \cap \mathrm{L}_T$ meets the dense set $\mathrm{D} \cap \mathrm{L}_T$, and so $\mathrm{V}$ meets $\mathrm{D}$.

(c). Pick $r \in \mathrm{R}$ and set $f(x) := x_r$, $x \in \mathrm{L}$. It follows from the properties of the Kurzweil-Stieltjes integral (Monteiro et al. 2018, Corollary 6.5.5(i), p. 172) that if $x \in \mathrm{L}$, then so does $f(x)$.

---

[9] As shown in Sokolov (2024), there is no nonconstant continuous real-valued function on $\mathrm{L}$ satisfying condition INT. That is, a profitability metric satisfying INT is necessarily partial (i.e., is undefined for some cash flow streams). In particular, so are the profitability index and the (discounted) payback period.

[10] For instance, if $E$ in condition (i) in the definition of an $E$-cap is the profitability index defined on $\{x \in \mathrm{L} : x(0) < 0\}$ by $E(x) = F(x)/(-x(0))$, $F \in \mathcal{NPV}$, then conditions (i)–(vi) are consistent and thus provide a meaningful concept of $E$-cap. An example of such an $E$-cap is given by $r \mapsto \{x \in \mathrm{L} : x(0) \le 0,\ F(x) + rx(0) \le 0\}$.

Clearly, $f : \mathrm{L} \to \mathrm{L}$ is linear and bijective. Since $f^{-1}(x) = x_{-r}$, we only have to show that $f$ is continuous. Using the estimate of the Kurzweil-Stieltjes integral (Monteiro et al. 2018, Theorem 6.3.7, p. 154), we get that for any $T = 1,2,...$, there is a constant $c > 0$ (which may depend on $r$ and $T$) such that $\|f(x) - x(0)1_0\| \le c\|x\|$ $\forall x \in \mathrm{L}_T$. Since $\|f(x)\| \le \|f(x) - x(0)1_0\| + \|x(0)1_0\| \le c\|x\| + \|x\| = (c+1)\|x\|$, this proves that for each $T = 1,2,...$, the restriction of $f$ to $\mathrm{L}_T$ is continuous and therefore so is $f$ (Narici and Beckenstein 2010, Theorem 12.2.2, p. 434).

(d). Follows from part (c) and the definition of the weak$^*$ topology.

(e). Pick $x \in \mathrm{L}$. There is $T$ such that $x_r \in \mathrm{L}_T$ for all $r \in \mathrm{R}$. In view of part (a), it is sufficient to prove that $r \mapsto x_r$ is continuous as a function from $\mathrm{R}$ to $\mathrm{L}_T$. For any $r \in \mathrm{R}$, denote by $h_r$ the function on $[0,T]$ given by $h_r(t) := e^{-rt}$. Note that $h_r$ converges pointwise to $h_s$ as $r \to s$ and $h_r \ge h_s$ (pointwise) whenever $r < s$. Therefore, by the Dini theorem (Aliprantis and Border 2006, Theorem 2.66, p. 54), $h_r$ converges uniformly to $h_s$ as $r \to s$. Pick $r,s \in \mathrm{R}$. For any $t \in [0,T]$, using the standard estimate of the Kurzweil-Stieltjes integral (Monteiro et al. 2018, Theorem 6.3.6, p. 154), we get $|x_r(t) - x_s(t)| \le V_0^t(x)\|h_r - h_s\|_0^t \le V_0^T(x)\|h_r - h_s\|_0^T$ and, therefore, $\|x_r - x_s\|_0^T \le V_0^T(x)\|h_r - h_s\|_0^T$. Since $h_r$ converges uniformly to $h_s$ as $r \to s$, we are done. □

**Lemma 4.**

*$F : \mathrm{L} \to \mathrm{R}$ is an NPV functional, i.e., $F \in \mathcal{NPV}$, if and only if representation (1) holds for some $\alpha \in \mathcal{A}$.*

**Proof.**

Assume that Eq. (1) holds for some $\alpha \in \mathcal{A}$. Since $F$ is linear and its restriction to each $\mathrm{L}_T$, $T = 1,2,...$ is continuous (Monteiro et al. 2018, Theorem 8.2.8, p. 304), we have $F \in \mathrm{L}^*$ (Narici and Beckenstein 2010, Theorem 12.2.2, p. 434). Clearly, $F(1_0) = 1$. So we only have to show that $F(x) \le 0$ for all $x \in \mathrm{L}_-$. Pick $x \in \mathrm{L}_-$. There is $T$ such that $x \in \mathrm{L}_T$. Since $\mathrm{D} \cap \mathrm{L}_T$ is dense in $\mathrm{L}_T$ (Lemma 3, part (b)), for any $\varepsilon > 0$, there is a loan $y = \sum_{k=1}^{n} c_k 1_{t_k} \in \mathrm{D} \cap \mathrm{L}_T$, $c_1,...,c_n \in \mathrm{R}$, $0 \le t_1 < ... < t_n$ such that $\|x - y\| < \varepsilon$. The constants $c_1,...,c_n$ can be chosen such that $y \in \mathrm{L}_-$, i.e., $c_1 + ... + c_k \le 0$ for all $k = 1,...,n$. Indeed, the loan $y_-(t) := \min\{y(t),0\}$ satisfies $y_- \in \mathrm{L}_-$ and $\|x - y_-\| < \varepsilon$. As $\alpha$ is nonnegative and nonincreasing, we have

$$F(y) = \sum_{k=1}^{n} c_k \alpha(t_k) = \alpha(t_n)(c_1 + ... + c_n) + \sum_{k=1}^{n-1} (\alpha(t_k) - \alpha(t_{k+1}))(c_1 + ... + c_k) \le 0 .$$

Combining this with part (a) of Lemma 3, we get that for any neighborhood $\mathrm{U}$ of $x$ there is $y \in \mathrm{U}$ such that $F(y) \le 0$. Since $F$ is continuous, this proves that $F(x) \le 0$.

Now assume that $F \in \mathcal{NPV}$, i.e., $F \in \mathrm{L}_-^\circ$ and $F(1_0)=1$. Let $\alpha : \mathrm{R}_+ \to \mathrm{R}$ be the function defined by $\alpha(t) := F(1_t)$. $\alpha$ is nonnegative: for any $t \in \mathrm{R}_+$, we have $-1_t \in \mathrm{L}_-$ and therefore $\alpha(t)=F(1_t)=-F(-1_t)\ge 0$. $\alpha$ is nonincreasing: for any $t<\tau$, we have $\alpha(t)-\alpha(\tau)=F(1_t)-F(1_\tau)=-F(-1_t+1_\tau)\ge 0$ as $-1_t+1_\tau \in \mathrm{L}_-$. Since $\alpha(0)=F(1_0)=1$, we get $\alpha \in \mathcal{A}$. We have to show that for each $T$, representation (1) holds for all $x \in \mathrm{L}_T$. Pick $T$ and note that $\mathrm{L}_T$ is homeomorphic to the space of restrictions of elements of $\mathrm{L}_T$ to the set $[0,T]$ endowed with the topology of uniform convergence. Since $F \in \mathrm{L}^*$, the restriction of $F$ to $\mathrm{L}_T$ is an element of $\mathrm{L}_T^*$. Therefore, there exists a function of bounded variation $\alpha_T : [0,T] \to \mathrm{R}$ such that $F(x)=\alpha_T(0)x(0)+\int_0^T \alpha_T(t)\mathrm{d}x(t)$ for all $x \in \mathrm{L}_T$ (Monteiro et al. 2018, Theorem 8.2.8, p. 304). As $\alpha_T(t)=F(1_t)$, $t\in[0,T]$, we obtain that $\alpha_T$ is the restriction of $\alpha$ to $[0,T]$. Hence, representation (1) holds for all $x \in \mathrm{L}_T$. □

Part (b) of the next lemma generalizes the results of Gronchi (1986, Proposition 1), Hazen (2003, Theorem 3), and Promislow (2015, Theorem 2.1, p. 29).

**Lemma 5.**

*The following statements hold.*

(a) *For all* $x \in \mathrm{L}$ *and* $r \le s$, $x_r \in \mathrm{L}_- \;\Rightarrow\; x_s \in \mathrm{L}_-$.

(b) $\mathrm{S}_1 \subset \mathrm{S}_2$, *where* $\mathrm{S}_1$ *and* $\mathrm{S}_2$ *are defined in (2).*

**Proof.**

We shall prove only part (b). Let $x \in \mathrm{S}_1$, i.e., $x \in \mathrm{L}\setminus\{0_{\mathrm{L}}\}$ and there is $\lambda \in \mathrm{R}$ such that $x_\lambda$ is nonpositive and $x_\lambda(+\infty)=0$. Let $T$ be the maturity date of $x$. For any $r \in \mathrm{R}$, using integration by substitution and integration by parts (Monteiro et al. 2018, Theorems 6.4.2 and 6.6.1, pp. 167, 175, 176), we get

$$
\begin{aligned}
x_{\lambda+r}(+\infty) &= x(0)+\int_0^T e^{-(\lambda+r)t}\mathrm{d}x(t) = x(0)+\int_0^T e^{-rt}e^{-\lambda t}\mathrm{d}x(t) = x(0)+\int_0^T e^{-rt}\mathrm{d}\big(x_\lambda(t)-x(0)\big) \\
&= x(0)+\int_0^T e^{-rt}\mathrm{d}x_\lambda(t) = x(0)+e^{-rT}x_\lambda(T)-x_\lambda(0)-\int_0^T x_\lambda(t)\mathrm{d}(e^{-rt}) = r\int_0^T x_\lambda(t)e^{-rt}\mathrm{d}t\,,
\end{aligned}
\tag{8}
$$

where we use that $x_\lambda(0)=x(0)$ and $x_\lambda(T)=x_\lambda(+\infty)=0$. Since $x_\lambda$ is nonzero, nonpositive, right-continuous, and therefore negative on a nondegenerate interval, it follows from (8) that $F_{\lambda+r}(x)=x_{\lambda+r}(+\infty)\lessgtr 0$ whenever $r \gtrless 0$. □

**Proof of Proposition 1.**

We shall prove part (b) and give only a sketch of a proof of part (a).

(b). *Claim 1*: if $\mathrm{S}_0 \subseteq \mathrm{P} \subseteq \mathrm{S}_3$, then the restriction of $I_3$ to $\mathrm{P}$ is a continuous IRR on $\mathrm{P}$.

Let $\mathrm{S}_0 \subseteq \mathrm{P} \subseteq \mathrm{S}_3$ and $E:\mathrm{P} \to \mathrm{R}$ be the restriction of $I_3$ to $\mathrm{P}$. Clearly, $E$ is an IRR on $\mathrm{P}$. To show that $E$ is continuous, note that for any $r \in \mathrm{R}$, $\{x \in \mathrm{P} : E(x) \le r\}$ is closed in $\mathrm{P}$ as the intersection of a closed in $\mathrm{L}$ set $\{F_s,\, s \in [r,+\infty)\}^\circ$ and $\mathrm{P}$. Similarly, for any $r \in \mathrm{R}$, $\{x \in \mathrm{P} : E(x) \ge r\}$ is closed in $\mathrm{P}$ as the intersection of a closed in $\mathrm{L}$ set $\{-F_s,\, s \in (-\infty, r]\}^\circ$ and $\mathrm{P}$.

*Claim 2*: for any continuous IRR on a superset of $\mathrm{S}_1$, its restriction to $\mathrm{S}_1$ is $I_1$.

Let $E$ be a continuous IRR on a superset of $\mathrm{S}_1$. First, we show that $x \in \mathrm{S}_1$ & $I_1(x) = 0 \Rightarrow E(x) = 0$. Pick $x \in \mathrm{S}_1$ with $I_1(x) = 0$. There is $T$ such that $x \in \mathrm{L}_T$. Since $\mathrm{D} \cap \mathrm{L}_T$ is dense in $\mathrm{L}_T$ (Lemma 3, part (b)), for any $\varepsilon > 0$, there is $y = \sum_{k=1}^{n+1} c_k 1_{t_k} \in \mathrm{D} \cap \mathrm{L}_T$, $c_1,\dots,c_{n+1} \in \mathrm{R}$, $0 \le t_1 < \dots < t_{n+1}$ such that $\|x - y\| < \varepsilon$. As $x \in \mathrm{S}_1$ and $I_1(x) = 0$, the constants $c_1,\dots,c_{n+1}$ can be chosen such that $c_1 + \dots + c_k < 0$ for all $k = 1,\dots,n$ and $c_1 + \dots + c_{n+1} = 0$. In this case, $y \in \mathrm{S}_1$ and $I_1(y) = 0$. Set $y^{(k)} := (c_1 + \dots + c_k)(1_{t_k} - 1_{t_{k+1}})$, $k = 1,\dots,n$. Then for each $k = 1,\dots,n$, we have $y^{(k)} \in \mathrm{S}_0$, $E(y^{(k)}) = I_0(y^{(k)}) = 0$ (by CONS), and $\sum_{i=1}^{k} y^{(i)} \in \mathrm{S}_1$. As

$$\sum_{k=1}^{n} y^{(k)} = \sum_{k=1}^{n} (c_1 + \dots + c_k)(1_{t_k} - 1_{t_{k+1}}) = \sum_{k=1}^{n} c_k (1_{t_k} - 1_{t_{n+1}}) = \sum_{k=1}^{n} c_k 1_{t_k} - \left( \sum_{k=1}^{n} c_k \right) 1_{t_{n+1}} = \sum_{k=1}^{n+1} c_k 1_{t_k} = y,$$

condition INT implies $E(y) = E\left( \sum_{k=1}^{n} y^{(k)} \right) = 0$. Combining this with Lemma 3 (part (a)), we get that for any neighborhood $\mathrm{U}$ of $x$ there is $y \in \mathrm{U} \cap \mathrm{S}_1$ such that $E(y) = 0$. Since $E$ is continuous, this proves that $E(x) = 0$.

Now pick $r \in \mathrm{R}$ and notice the following three facts: (I) for any $T = 1,2,\dots$, the map $x \mapsto x_r$ is a linear self-homeomorphism of $\mathrm{L}_T$ (Lemma 3, part (c)) and, therefore, there are $m, M \in \mathrm{R}_{++}$ such that $m\|x\| \le \|x_r\| \le M\|x\|$ for every $x \in \mathrm{L}_T$; (II) $x \mapsto x_r$ maps $\mathrm{S}_1 \cap \mathrm{L}_T$ and $\mathrm{D} \cap \mathrm{L}_T$, $T = 1,2,\dots$ onto themselves; (III) for any $x \in \mathrm{S}_1$, $I_1(x) = r \Leftrightarrow I_1(x_r) = 0$. Using these facts and reproducing the above proof of "$x \in \mathrm{S}_1$ & $I_1(x) = 0 \Rightarrow E(x) = 0$" with $x$ replaced by $x_r$, we obtain that $x \in \mathrm{S}_1$ & $I_1(x) = r \Rightarrow E(x) = r$.

*Claim 3*: if $\mathrm{S}_1 \subseteq \mathrm{P} \subseteq \mathrm{S}_3$, then the restriction of $I_3$ to $\mathrm{P}$ is a unique continuous IRR on $\mathrm{P}$.

Let $\mathrm{S}_1 \subseteq \mathrm{P} \subseteq \mathrm{S}_3$ and $E$ be a continuous IRR on $\mathrm{P}$. Pick $x \in \mathrm{P}$. The function $s \mapsto F_s(x)$ is nonzero and real analytic (Widder 1946, Theorem 5a, p. 57), so the set of its zeros is nowhere dense in $\mathrm{R}$ (Krantz and Parks 2002, Corollary 1.2.7, p. 14). Therefore, for any $\varepsilon > 0$, there is $r \in (I_3(x) - \varepsilon, I_3(x))$ such that $F_r(x) > 0$. Set $y := -c1_0 + (c - x_r(T))e^{rT} 1_T$, where $T$ is the maturity date of $x$ and $c > 0$. If $c$ is large enough, then $y \in \mathrm{S}_0$, $x + y \in \mathrm{S}_1$, and $I_1(x + y) = r$. By Claim 2, we have $E(x + y) = I_1(x + y) = r$. On the other hand, since $x_r(T) = F_r(x) > 0$, from CONS it follows that $E(y) = I_0(y) < r$. As $E(y) < E(x + y) = r$, condition INT implies $E(x) \ge r$. Since $\varepsilon > 0$ is arbitrary, this proves that $E(x) \ge I_3(x)$. A similar argument shows that $E(x) \le I_3(x)$.

*Claim 4*: if $S_1 \subseteq P \subseteq L$ and $P \setminus S_3 \neq \varnothing$, then there is no continuous IRR on $P$.

Let $S_1 \subseteq P \subseteq L$ and $P \setminus S_3 \neq \varnothing$. Assume by way of contradiction that $E$ is a continuous IRR on $P$. Pick $x \in P \setminus S_3$ and set $f(r) := F_r(x)$. We consider four cases.

CASE 1: $x = 0_L$. Pick $y \in S_0$ with $I_0(y) \neq E(0_L)$. Using CONS and continuity of $E$, we get $I_0(y) = \lim_{\lambda \to 0+} I_0(\lambda y) = \lim_{\lambda \to 0+} E(\lambda y) = E(0_L)$, which is a contradiction.

CASE 2: $x \neq 0_L$ and $f$ is nonpositive. As $f$ is nonzero and real analytic, there is $r < E(x)$ such that $f(r) < 0$. Set $y := -c1_0 + (c - x_r(T))e^{rT} 1_T \in S_0$, where $T$ is the maturity date of $x$ and $c > 0$. Note that $(x+y)_r(+\infty) = (x+y)_r(T) = 0$ and $(x+y)_r \in L_-$ for sufficiently large $c$. Therefore, $x + y \in S_1$ with that $c$ and, by Claim 2, $E(x+y) = I_1(x+y) = r$. Since $x_r(T) = f(r) < 0$, we have $E(y) = I_0(y) > r$. As $r = E(x+y) < E(y)$, condition INT implies $E(x) \leq r$, which is a contradiction.

CASE 3: $x \neq 0_L$ and $f$ is nonnegative. In a similar manner as in Case 2, we arrive at a contradiction.

CASE 4: there are $r_1 < r_2$ such that $f(r_1) < 0 < f(r_2)$. As in Case 2, one can show that there are $y^{(1)}, y^{(2)} \in S_0$ such that $E(y^{(1)}) > r_1$, $E(y^{(2)}) < r_2$, $x + y^{(i)} \in S_1$, and $E(x + y^{(i)}) = r_i$, $i = 1,2$. As $E(y^{(1)}) > E(x + y^{(1)}) = r_1$ and $E(y^{(2)}) < E(x + y^{(2)}) = r_2$, condition INT implies $E(x) \leq r_1$ and $E(x) \geq r_2$, which is a contradiction.

(a). Clearly, the restriction of $J_3$ to a set $P$, where $S_0 \subseteq P \subseteq D_3$, is an IRR on $P$.

A similar argument to that used to prove Claim 2 in part (b) shows that if $x \in D_1$ and $J_1(x) = r$, then there are $x^{(k)} \in S_0$, $k = 1, \ldots, n$ such that $I_0(x^{(k)}) = r$, $\sum_{i=1}^{k} x^{(i)} \in D_1$, and $\sum_{k=1}^{n} x^{(k)} = x$.[11] With the help of conditions CONS and INT, this proves that for any IRR on a superset of $D_1$, its restriction to $D_1$ is $J_1$. Using this result and reproducing the proof of Claim 3 in part (b) with $S_1$ and $S_3$ replaced by $D_1$ and $D_3$, we get that any IRR on a set $P$, where $D_1 \subseteq P \subseteq D_3$, is the restriction of $J_3$ to $P$. Finally, similar arguments to those used in the proofs of Cases 2–4 in Claim 4 in part (b) show that if $D_1 \subseteq P \subseteq D \setminus \{0_L\}$ and $P \setminus D_3 \neq \varnothing$, then there is no IRR on $P$. □

**Proof of Proposition 2.**

We shall prove only part (b); the argument for part (a) is similar.

(b). *Claim 1*: if $S_1 \subseteq P \subseteq S_2$, then the restriction of $I_2$ to $P$ is a continuous strict IRR on $P$.

Let $S_0 \subseteq P \subseteq S_2$ and $E$ be the restriction of $I_2$ to $P$. Being the restriction of $I_3$, $E$ is a continuous IRR on $P$. Clearly, it satisfies S-INT.

[11] See Proposition 2 in Gronchi (1986) for a related result.

*Claim 2*: if $\mathrm{S}_1 \subseteq \mathrm{P} \subseteq \mathrm{L}$ and $E$ is a continuous strict IRR on $\mathrm{P}$, then $\mathrm{P} \subseteq \mathrm{S}_3$ and $E$ is the restriction of $I_3$ to $\mathrm{P}$.

This follows from part (b) of Proposition 1.

*Claim 3*: if $\mathrm{S}_1 \subseteq \mathrm{P} \subseteq \mathrm{L}$ and $\mathrm{P} \setminus \mathrm{S}_2 \neq \varnothing$, then there is no continuous strict IRR on $\mathrm{P}$.

Let $\mathrm{S}_1 \subseteq \mathrm{P} \subseteq \mathrm{L}$ and $\mathrm{P} \setminus \mathrm{S}_2 \neq \varnothing$. Assume by way of contradiction that $E$ is a continuous strict IRR on $\mathrm{P}$. Pick $x \in \mathrm{P} \setminus \mathrm{S}_2$. By Claim 2, $x \in \mathrm{S}_3 \setminus \mathrm{S}_2$ and, therefore, the function $r \mapsto F_r(x)$ has at least two zeros. Let $r_1$, $r_2$ be distinct zeros of $r \mapsto F_r(x)$. Set $y^{(i)} = -c1_0 + ce^{r_i T} 1_T$, $i = 1,2$, where $T$ is the maturity date of $x$ and $c > 0$. Then $y^{(i)} \in \mathrm{S}_0$, $E(y^{(i)}) = I_0(y^{(i)}) = r_i$ (by CONS), and, for sufficiently large $c$, $x + y^{(i)} \in \mathrm{S}_1$, $i = 1,2$. As $F_{r_i}(x) = F_{r_i}(y^{(i)}) = 0$, we have $F_{r_i}(x + y^{(i)}) = 0$ and therefore $I_1(x + y^{(i)}) = r_i$, $i = 1,2$. By Claim 2, $E$ reduces to $I_1$ on $\mathrm{S}_1$, so $E(x + y^{(i)}) = I_1(x + y^{(i)}) = r_i$, $i = 1,2$. Since $E(y^{(i)}) = E(x + y^{(i)}) = r_i$, condition S-INT implies $E(x) = r_i$, $i = 1,2$, which is a contradiction. □

**Lemma 6.**

*For $\alpha \in \mathcal{A}$ and $r \in \mathrm{R}$, the following conditions are equivalent.*

(a) *The function $t \mapsto e^{rt}\alpha(t)$ is nonincreasing.*

(b) $-(\ln \alpha(t))' \geq r$, $t \in \mathrm{R}_+$, *whenever the left-hand side of the inequality is defined.*

**Proof.**

(a)$\Rightarrow$(b). Follows by taking logs of $t \mapsto e^{rt}\alpha(t)$ and differentiating.

(b)$\Rightarrow$(a). Pick $0 \leq s < t$. If $\alpha(t) = 0$, then $e^{rs}\alpha(s) \geq e^{rt}\alpha(t)$ holds trivially. Now assume that $\alpha(t) > 0$. As $\alpha \in \mathcal{A}$, the function $\tau \mapsto -\ln \alpha(\tau)$ is nondecreasing on $[s,t]$. We have $\ln \alpha(s) - \ln \alpha(t) \geq \int_s^t (-\ln \alpha(\tau))' \mathrm{d}\tau \geq \int_s^t r \mathrm{d}\tau = r(t-s)$, where the first inequality follows from a result on Lebesgue integrability of the derivative of a nondecreasing function (Kadets, 2018, Theorem 1, p. 191) and the second inequality follows from (b). Rearranging the terms and taking the exponential of both sides, we get $e^{rs}\alpha(s) \geq e^{rt}\alpha(t)$. □

**Proof of Proposition 7.**

(a)$\Rightarrow$(b). Trivial.

(b)$\Rightarrow$(c). Let $\mathcal{N}$ be a weak cap. Pick $r \in \mathrm{R}_+$ and set $\mathcal{F}(r) := \mathcal{N}(r)^\circ \cap \mathcal{NPV}$.

*Claim 1*: $\mathcal{F}(r)^\circ = \mathcal{N}(r)$.

Conditions (iv), (v), and (vi) in the definition of a weak cap imply that $\mathcal{N}(r)$ is a closed convex cone. In particular, $0_\mathrm{L} \in \mathcal{N}(r)$. Condition (ii) with $y = 0_\mathrm{L}$ implies $\mathrm{L}_- \subseteq \mathcal{N}(r)$; thus, $\mathcal{N}(r)^\circ \subseteq \mathrm{L}_-^\circ$. Since $-1_0$ is an interior point of $\mathrm{L}_-$, for each nonzero $F \in \mathrm{L}_-^\circ$, we have $F(1_0) > 0$ (Aliprantis and Border 2006, Lemma 5.66, p. 202). This proves that $\mathcal{NPV}$ is a base for the cone $\mathrm{L}_-^\circ$ (i.e., for each nonzero functional $F \in \mathrm{L}_-^\circ$, there is a unique $\lambda \in \mathrm{R}_{++}$ such that $\lambda F \in \mathcal{NPV}$). As $\mathcal{N}(r)^\circ \subseteq \mathrm{L}_-^\circ$, the set $\mathcal{N}(r)^\circ \cap \mathcal{NPV}$ is a base for the cone $\mathcal{N}(r)^\circ$. We have $\mathcal{F}(r)^\circ = (\mathcal{N}(r)^\circ \cap \mathcal{NPV})^\circ = (\mathcal{N}(r)^\circ)^\circ = \mathcal{N}(r)$, where the second equality follows from the fact that $\mathcal{N}(r)^\circ \cap \mathcal{NPV}$ is a base for the cone $\mathcal{N}(r)^\circ$ and the last equality follows from the bipolar theorem (Aliprantis and Border 2006, Theorem 5.103, p. 217).

*Claim 2*: if $F^{(\alpha)} \in \mathcal{F}(r)$, then $-(\ln \alpha(t))' \geq r$, $t \in \mathrm{R}_+$, whenever the left-hand side of the inequality is defined.

Pick $F^{(\alpha)} \in \mathcal{F}(r)$. Choose $0 \leq s < t$ and set $x = -e^{rs}1_s + e^{rt}1_t$. Note that $x \in \mathrm{S}_0$ and $I_0(x) = r$. Condition (i) implies that $x \in \mathcal{N}(r)$. We have $0 \geq F^{(\alpha)}(x) = -e^{rs}\alpha(s) + e^{rt}\alpha(t)$, where the inequality follows from Claim 1. This proves that the function $t \mapsto e^{rt}\alpha(t)$ is nonincreasing. Combining this with Lemma 6, the claim follows.

*Claim 3*: for any $T \geq 0$ and $\delta > 0$, there is $F^{(\alpha)} \in \mathcal{F}(r)$ such that $1 - e^{rT}\alpha(T) < \delta$.

If $T = 0$, the claim holds trivially. So in what follows, we assume that $T > 0$.Without loss of generality we may assume that $\delta \in (0,1)$ and consider $x = -(1-\delta)1_0 + e^{rT}1_T \in \mathrm{S}_0$. Since $I_0(x) > r$, condition (i) in the definition of a weak cap implies $x \notin \mathcal{N}(r)$. That is, there is $F^{(\alpha)} \in \mathcal{F}(r)$ such that $0 < F^{(\alpha)}(x) = -(1-\delta) + e^{rT}\alpha(T)$ and the claim follows.

*Claim 4*: $F_r \in \mathcal{F}(r)$.

By the definition of the weak* topology, for any neighborhood $\mathrm{U}$ of $F_r$ in $\mathcal{NPV}$, there are $x^{(1)},\ldots,x^{(n)} \in \mathrm{L}$ and $\varepsilon > 0$ such that $\{F \in \mathcal{NPV} : |F_r(x^{(i)}) - F(x^{(i)})| < \varepsilon, i = 1,\ldots,n\} \subseteq \mathrm{U}$. By construction, the set $\mathcal{F}(r)$ is closed in $\mathcal{NPV}$. Therefore, to prove the claim, it suffices to show that for any $x^{(1)},\ldots,x^{(n)} \in \mathrm{L}$ and $\varepsilon > 0$, there is $F \in \mathcal{F}(r)$ such that $|F_r(x^{(i)}) - F(x^{(i)})| < \varepsilon$, $i = 1,\ldots,n$.

Pick $x^{(1)},\ldots,x^{(n)} \in \mathrm{L}$, $\varepsilon > 0$ and set $T := \max_{i \in \{1,\ldots,n\}} T_i$, where $T_i$ is the maturity date of $x^{(i)}$, $i = 1,\ldots,n$. There is $F^{(\alpha)} \in \mathcal{F}(r)$ such that $2(1 - e^{rT}\alpha(T))\|x^{(i)}\| < \varepsilon$ for all $i = 1,\ldots,n$ (if $\max_{i \in \{1,\ldots,n\}} \|x^{(i)}\| \neq 0$, then this follows from Claim 3 with $\delta = \varepsilon / (2 \max_{i \in \{1,\ldots,n\}} \|x^{(i)}\|)$; otherwise it holds trivially). Put $f(t) := e^{-rt}$, $g(t) := 1 - e^{rt}\alpha(t)$ with that $\alpha$ and note that $g$ is nondecreasing (Claim 2 and Lemma 6) and $g(0) = 0$. We have

$$|F_r(x^{(i)}) - F^{(\alpha)}(x^{(i)})| = \left|\int_0^T (e^{-rt} - \alpha(t))\mathrm{d}x^{(i)}(t)\right| = \left|\int_0^T f(t)g(t)\mathrm{d}x^{(i)}(t)\right|$$

$$\leq \left[|f(0)g(0)| + |f(T)g(T)| + V_0^T(fg)\right]\|x^{(i)}\| \leq \left[|f(T)g(T)| + V_0^T(f)\|g\|_0^T + V_0^T(g)\|f\|_0^T\right]\|x^{(i)}\|$$

$$= \left[e^{-rT}(1 - e^{rT}\alpha(T)) + (1 - e^{-rT})(1 - e^{rT}\alpha(T)) + (1 - e^{rT}\alpha(T))\right]\|x^{(i)}\| = 2(1 - e^{rT}\alpha(T))\|x^{(i)}\| < \varepsilon,\ i = 1,\ldots,n,$$

where the first inequality follows from the estimate of the Kurzweil-Stieltjes integral (Monteiro et al. 2018, Theorem 6.3.7, p. 154) and the second inequality follows from the estimate of the total variation of the product of functions of bounded variations.

(c)$\Rightarrow$(d). Pick $r \in \mathrm{R}_+$. Condition 1 in part (c) implies that $\mathcal{N}(r)$ is a closed convex cone.

The least (by inclusion) subset $\mathcal{F}(r)$ of $\mathcal{NPV}$ satisfying conditions 2 and 3 is $\{F_r\}$. This proves the second inclusion in (6).

Let us prove the first inclusion in (6). It follows from Lemma 6 that the greatest (by inclusion) subset $\mathcal{F}(r)$ of $\mathcal{NPV}$ satisfying conditions 2 and 3 is $\mathcal{H}(r) := \{ F^{(\alpha)} \in \mathcal{NPV} :\ t \mapsto e^{rt}\alpha(t)$ is nonincreasing$\}$. Note that $\mathcal{H}(0) = \mathcal{NPV}$. It is straightforward to verify that $x \in \mathcal{H}(r)^\circ \iff x_r \in \mathcal{H}(0)^\circ$. Since $\mathcal{H}(0)^\circ = \mathcal{NPV}^\circ = \mathrm{L}_-$, we are done.

(d)$\Rightarrow$(a). Pick $r \in \mathrm{R}_+$. Since $\mathcal{N}(r)$ is a closed convex cone, conditions (iv), (v), and (vi) in the definition of a weak $E$-cap hold. As $\mathrm{L}_- \subseteq \mathcal{N}_-(r) \subseteq \mathcal{N}(r)$ (where the first inclusion follows from part (a) of Lemma 5) and $\mathcal{N}(r)$ is a convex cone, condition (ii) holds.

Finally, to verify condition (i), pick $x \in \mathrm{P}$ and set $s = E(x) = I_1(x)$. If $s \le r$, then $\mathcal{N}_-(s) \subseteq \mathcal{N}_-(r) \subseteq \mathcal{N}(r)$, where the first inclusion follows from part (a) of Lemma 5 and the second inclusion follows from (6). As $x \in \mathcal{N}_-(s)$ (by the definition of $I_1(x)$), we have $x \in \mathcal{N}(r)$. Now assume that $s > r$. Since $s = I_1(x) = I_2(x)$, we have $F_r(x) > 0$, or in other words $x \notin \{F_r\}^\circ$. As $\mathcal{N}(r) \subseteq \{F_r\}^\circ$, we get $x \notin \mathcal{N}(r)$. □

**Proof of Proposition 4.**

(a)$\Rightarrow$(b). Trivial.

(b)$\Rightarrow$(c). Since each cap is a weak cap, by Proposition 7, there is a correspondence $\mathcal{F} : \mathrm{R}_+ \rightrightarrows \mathcal{NPV}$ such that for any $r \in \mathrm{R}_+$, properties 1, 2, and 3 in part (c) hold. It follows from the proof of Proposition 7 that $\mathcal{F}$ can be chosen to be $\mathcal{F}(r) = \mathcal{N}(r)^\circ \cap \mathcal{NPV}$ $\forall r \in \mathrm{R}_+$. In this case, condition (iii) in the definition of a cap implies property 4.

(c)$\Rightarrow$(d). Conditions 1 and 4 in part (c) imply (iii)–(vi). The least (by inclusion) subsets $\mathcal{F}(r)$, $r \in \mathrm{R}_+$ of $\mathcal{NPV}$ satisfying conditions 2, 3, and 4 in part (c) are $\{F_s,\ s \in [r, +\infty)\}$, $r \in \mathrm{R}_+$. This proves the second inclusion in (4). The first inclusion in (4) follows from the one in (6) and the fact that $\mathcal{N}_-(r) \subseteq \mathcal{N}_-(s)$ for any $r \le s$ (Lemma 5, part (a)).

(d)$\Rightarrow$(a). Follows from the implication "(d)$\Rightarrow$(a)" in Proposition 7. □

**Proof of Proposition 6.**

(a)$\Rightarrow$(b). It follows from Proposition 7 that for any $r \in \mathrm{R}_+$, $\mathcal{N}(r)$ is a closed convex cone satisfying $\mathcal{N}(r) \subseteq \{F_r\}^\circ$. So we need only to show that for any $r \in \mathrm{R}_+$, $\mathcal{N}_+(r) \subseteq \mathcal{N}(r)$.

Let $\mathcal{C}(r)$, $r \in \mathrm{R}_+$ be the closed convex hull of $\{F_s,\ s \in [r, +\infty)\}$. Since the set $\mathcal{NPV}$ is closed and convex, $\mathcal{C}(r) \subset \mathcal{NPV}$. By Proposition 7, there is a correspondence $\mathcal{F} : \mathrm{R}_+ \rightrightarrows \mathcal{NPV}$ such that $\mathcal{N}(r) = \mathcal{F}(r)^\circ$. We have to show that for any $r \in \mathrm{R}_+$, $\mathcal{F}(r) \subseteq \mathcal{C}(r)$.

Let us first prove that $\mathcal{F}(0)\subseteq\mathcal{C}(0)$. Assume by way of contradiction that $\mathcal{F}(0)\setminus\mathcal{C}(0)\neq\varnothing$ and pick some $H\in\mathcal{F}(0)\setminus\mathcal{C}(0)$. To arrive at a contradiction, it is sufficient to show that there is $z\in\mathrm{D}_2$ such that $J_2(z)=0$ and $H(z)>0$ (indeed, as $\mathrm{D}_2\subseteq\mathrm{P}$ and $J_2(z)=0$, condition (i) implies $z\in\mathcal{N}(0)$; on the other hand, since $H(z)>0$, we have $z\notin\mathcal{N}(0)$, which is a contradiction). It follows from the bipolar theorem that there is $x\in\mathrm{L}$ such that $H(x)>0$ and $x\in\mathcal{C}(0)^{\circ}$. Since $H$ is continuous and $\mathrm{D}$ is dense in $\mathrm{L}$ (Lemma 3, part (b)), there exist $y\in\mathrm{D}$ and $\lambda>0$ such that $y+\lambda 1_0\le x$ and $H(y)>0$. Set $f(r):=F_r(y)$ and $g_\tau(r):=F_r(y-F_0(y)1_\tau)=f(r)-f(0)e^{-r\tau}$, $\tau\in\mathrm{R}_{++}$. As $y+\lambda 1_0\le x$ and $x\in\mathcal{C}(0)^{\circ}$, we have $f(r)\le-\lambda$ for all $r\in\mathrm{R}_+$.

*Claim 1*: $g_\tau$ is negative on $\mathrm{R}_{++}$ for sufficiently large $\tau>0$.

The claim holds trivially if $f(r)\le f(0)$ for all $r\in\mathrm{R}_{++}$ (i.e., $f$ attains its maximum on $\mathrm{R}_+$ at 0). Now assume that $f$ does not attain its maximum at 0. Note that in this case we have $f(0)<-\lambda$. Being real analytic, $f$ is continuously differentiable. In particular, the function $m(r):=\max\limits_{s\in[0,r]}f'(s)$ is well defined, continuous, and nondecreasing. Since $f$ does not attain its maximum at 0, $m(r)$ is positive for sufficiently large $r$, and there is $\bar r>0$ that solves $f(0)+m(\bar r)\bar r=-\lambda$. It follows from the definition of $m$ that $f(0)+m(\bar r)r\ge f(r)$ for all $r\in[0,\bar r]$. Thus, for any $\tau\in\mathrm{R}_+$, the function $h_\tau:\mathrm{R}_+\to\mathrm{R}$ defined by

$$h_\tau(r):=\begin{cases} f(0)+m(\bar r)r-f(0)e^{-r\tau} & \text{if } r\in[0,\bar r]\\ -\lambda-f(0)e^{-r\tau} & \text{if } r\in(\bar r,+\infty)\end{cases}$$

majorizes $g_\tau$ on $\mathrm{R}_+$. Note that $h_\tau(0)=0$, $h_\tau$ is convex on $r\in[0,\bar r]$ and decreasing on $r\in[\bar r,+\infty)$. Thus, $h_\tau$ is negative on $\mathrm{R}_{++}$ if and only if $h_\tau(\bar r)<0$. So $g_\tau$ is negative on $\mathrm{R}_{++}$ provided that $\tau$ satisfies the inequality $-\lambda-f(0)e^{-\bar r\tau}<0$.

*Claim 2*: $g_\tau$ is positive on $\mathrm{R}_{--}$ for sufficiently large $\tau>0$.

As $g_\tau(0)=0$, it is sufficient to prove that $g_\tau'$ is negative on $\mathrm{R}_-$ for sufficiently large $\tau$. First, let us show that there are $c,T\in\mathrm{R}_+$ such that $f'(r)\le ce^{-rT}$ for all $r\in\mathrm{R}_-$. Indeed, let $T$ be the maturity date of $y$. Using the standard estimate of the integral (Monteiro et al. 2018, Theorem 6.3.6, p. 154), we get

$$f'(r)=\int_0^T(-t)e^{-rt}\mathrm{d}y(t)\le V_0^T(y)\sup_{t\in[0,T]}\left|(-t)e^{-rt}\right|=V_0^T(y)Te^{-rT},\ r\in\mathrm{R}_-.$$

Setting $c=V_0^T(y)T$, we obtain $g_\tau'(r)=f'(r)-f(0)(-\tau)e^{-r\tau}\le ce^{-rT}+f(0)\tau e^{-r\tau}$, $r\in\mathrm{R}_-$. Since the right-hand side of the inequality is negative for $\tau>\max\{T,-c/f(0)\}$, Claim 2 holds.

Combining Claims 1 and 2, we get that there exists $\tau>0$ such that $g_\tau$ is positive on $\mathrm{R}_{--}$ and negative on $\mathrm{R}_{++}$. Set $z:=y-F_0(y)1_\tau$ with that $\tau$. As $F_r(z)=g_\tau(r)$, we have $z\in\mathrm{D}_2\subseteq\mathrm{P}$, $E(z)=J_2(z)=0$, and therefore, by condition (i), $z\in\mathcal{N}(0)$. On the other hand, since $F_0(y)<0$, $H(1_\tau)\ge0$, and, therefore, $H(z)=H(y)-F_0(y)H(1_\tau)\ge H(y)>0$, we get $z\notin\mathcal{N}(0)$, which is a contradiction. This proves that $\mathcal{F}(0)\subseteq\mathcal{C}(0)$.

Now we show that $\mathcal{F}(r)\subseteq\mathcal{C}(r)$ for all $r\in\mathrm{R}_{++}$. Pick $r\in\mathrm{R}_{++}$ and assume by way of contradiction that $\mathcal{F}(r)\setminus\mathcal{C}(r)\neq\varnothing$. Choose some $H\in\mathcal{F}(r)\setminus\mathcal{C}(r)$ and denote by $\overline{H}_{-r}$ the functional on $\mathrm{L}$ given by $x\mapsto H(x_{-r})$. As $H\in\mathcal{F}(r)$, from condition 3 in part (c) of Proposition 7 and Lemma 6 it follows that $\overline{H}_{-r}\in\mathcal{NPV}$. Since the map that sends each $F\in\mathrm{L}^*$ to the functional $x\mapsto F(x_r)$ is a linear self-homeomorphism of $\mathrm{L}^*$ (Lemma 3, part (d)), $F\in\mathcal{C}(0)$ if and only if the functional $x\mapsto F(x_r)$ belongs to $\mathcal{C}(r)$. Thus, $\overline{H}_{-r}\notin\mathcal{C}(0)$. Reproducing the proof of "$\mathcal{F}(0)\subseteq\mathcal{C}(0)$" with $H$ replaced by $\overline{H}_{-r}$, we get that there is $z\in\mathrm{D}_2$ such that $J_2(z)=0$ and $\overline{H}_{-r}(z)>0$. As $z_{-r}\in\mathrm{D}_2$ and $E(z_{-r})=J_2(z_{-r})=r$, by condition (i), we get $z_{-r}\in\mathcal{N}(r)$. On the other hand, since $H(z_{-r})=\overline{H}_{-r}(z)>0$, we have $z_{-r}\notin\mathcal{N}(r)$, which is a contradiction. This proves that $\mathcal{F}(r)\subseteq\mathcal{C}(r)$.

(b)$\Rightarrow$(a). It follows from Proposition 7 that $\mathcal{N}$ is a weak cap, so conditions (ii) and (iv)–(vi) hold. To verify condition (i), choose $r\in\mathrm{R}_+$, pick $x\in\mathrm{P}$ and set $s=E(x)=I_2(x)$. If $s\leq r$, then $x\in\mathcal{N}_+(r)$ (by the definition of $I_2(x)$). As $\mathcal{N}_+(r)\subseteq\mathcal{N}(r)$, we get $x\in\mathcal{N}(r)$. Now assume that $s>r$. By the definition of $I_2(x)$, we have $F_r(x)>0$, or in other words $x\notin\{F_r\}^\circ$. As $\mathcal{N}(r)\subseteq\{F_r\}^\circ$, we obtain $x\notin\mathcal{N}(r)$. □

**Proof of Proposition 3.**

(a)$\Rightarrow$(b). Being a weak $E$-cap, the correspondence $\mathcal{N}$ satisfies the conditions of part (b) of Proposition 6. Pick $r\in\mathrm{R}_+$. It follows from (5) that $\mathcal{N}_+(r)\subseteq\mathcal{N}(r)$. On the other hand, $\mathcal{N}(r)\subseteq\bigcap_{s:s\geq r}\mathcal{N}(s)\subseteq\bigcap_{s:s\geq r}\{F_s\}^\circ=\mathcal{N}_+(r)$, where the first inclusion follows from condition (iii) of the definition of an $E$-cap and the second inclusion follows from the second inclusion in (5). This proves that $\mathcal{N}(r)=\mathcal{N}_+(r)$.

(b)$\Leftrightarrow$(c). Pick $x\in\mathrm{L}$ and $r\in\mathrm{R}_+$.

*Claim 1*: $x\in\mathcal{N}_+(r)$ $\Rightarrow$ for any $y\in\mathrm{S}_2$ with $I_2(y)>r$, if $x+y\in\mathrm{S}_2$, then $I_2(x+y)\leq I_2(y)$.

Let $x\in\mathcal{N}_+(r)$ and $y\in\mathrm{S}_2$ be such that $I_2(y)>r$. As $F_s(x)\leq 0$ $\forall s\geq r$ and $F_s(y)<0$ $\forall s>I_2(y)$, we have $F_s(x+y)=F_s(x)+F_s(y)<0$ $\forall s>I_2(y)$. Therefore, if $x+y\in\mathrm{S}_2$, then $I_2(x+y)\leq I_2(y)$.

*Claim 2*: if for any $y\in\mathrm{S}_2$ with $I_2(y)>r$, $x+y\in\mathrm{S}_2$ implies $I_2(x+y)\leq I_2(y)$, then $x\in\mathcal{N}_+(r)$.

Let $x$ be such that for any $y\in\mathrm{S}_2$ with $I_2(y)>r$, if $x+y\in\mathrm{S}_2$, then $I_2(x+y)\leq I_2(y)$. Assume by way of contradiction that there is $s\in[r,+\infty)$ such that $F_s(x)>0$. Since the function $\lambda\mapsto F_\lambda(x)$ is continuous, without loss of generality, we may assume that $s>r$. To prove the claim, it is sufficient to show that there is $y\in\mathrm{S}_2$ such that $I_2(y)=s$ and $x+y\in\mathrm{S}_2$. Indeed, if there is such $y$, then $F_s(x+y)=F_s(x)+F_s(y)=F_s(x)>0$, and therefore, $I_2(x+y)>s=I_2(y)$, which is a contradiction.

Let $T$ be the maturity date of $x$. Set $y_c := c(-1_0 + e^{s(T+1)} 1_{T+1})$, $c > 0$. Clearly, for any $c > 0$, $y_c \in \mathrm{S}_2$ and $I_2(y_c) = s$. To complete the proof, we show that $x + y_c \in \mathrm{S}_2$ for sufficiently large $c$.

Denote $f(\lambda) := F_\lambda(x)$, $g_c(\lambda) := F_\lambda(y_c)$, $h_c(\lambda) := F_\lambda(x + y_c) = f(\lambda) + g_c(\lambda)$. Using the estimate $|f(\lambda)| = |x(0) + (f(\lambda) - x(0))| \le |x(0)| + 2\max\{1, e^{-\lambda T}\}\|x\|$ (Monteiro et al. 2018, Theorem 6.3.7, p. 154), we deduce that $h_c(-\infty) = +\infty$ $\forall c > 0$. The same estimate shows that $f$ is bounded on $\mathrm{R}_+$. Let $b \in \mathrm{R}$ be such that $f(\lambda) \le b$ for all $\lambda \in \mathrm{R}_+$. Choose arbitrary $\lambda^* > s$. Since $g_c$ is decreasing, for all $\lambda \in [\lambda^*, +\infty)$, $h_c(\lambda) = f(\lambda) + g_c(\lambda) \le f(\lambda) + g_c(\lambda^*) < b - c(1 - e^{-(\lambda^* - s)(T+1)})$. This estimate shows that $h_c$ is negative on $[\lambda^*, +\infty)$ for sufficiently large $c > 0$.

Using the estimate of the integral $f'(\lambda)$, we get that there is $d \ge 0$ such that $f'(\lambda) \le d e^{-\lambda T}$ for all $\lambda \in \mathrm{R}_-$. As $h_c'(\lambda) = f'(\lambda) - c(T+1) e^{s(T+1)} e^{-\lambda(T+1)}$, $h_c'$ is negative on $\mathrm{R}_-$ for sufficiently large $c$. Since $f$ is continuously differentiable, $f'$ is bounded on $[0, \lambda^*]$. Therefore, $h_c'$ is negative on $[0, \lambda^*]$ for sufficiently large $c$.

Summarizing, we get that there are $c > 0$ and $\lambda^* > s$ such that $h_c(-\infty) = +\infty$, $h_c$ is strictly decreasing on $(-\infty, \lambda^*]$ and negative on $[\lambda^*, +\infty)$. Thus, $x + y_c \in \mathrm{S}_2$ with that $c$.

(b)$\Rightarrow$(a). Straightforward. □

**Proof of Lemma 1.**

(a)$\Leftrightarrow$(b). This follows from parts (c) and (d) of Proposition 4 and the observation that the sets $\mathcal{F}(r) = \{F^{(\alpha)} \in \mathcal{NPV}\colon -(\ln \alpha(t))' \ge r,\ t \in \mathrm{R}_+$, whenever the left-hand side of the inequality is defined$\}$, $r \in \mathrm{R}_+$ are the largest (by inclusion) subsets of $\mathcal{NPV}$ satisfying conditions 2–4 in part (c) of Proposition 4.

(a)$\Rightarrow$(c). Let $x \in \mathcal{N}_(r)$. If $x = 0_{\mathrm{L}}$, then there is nothing to prove. Assume that $x \ne 0_{\mathrm{L}}$ and set $y := x - x_r(T) e^{r(T+1)} 1_{T+1}$, where $T$ is the maturity date of $x$. As $x \in \mathcal{N}_(r)$, we have $x_r(T) \le 0$. So $x \le y$, $y \in \mathrm{S}_1$, and $I_1(y) = r$.

(c)$\Rightarrow$(a). If $x = 0_{\mathrm{L}}$, then (a) holds trivially. Now suppose that $x \ne 0_{\mathrm{L}}$ and there exists $y \in \mathrm{S}_1$ such that $x \le y$ and $I_1(y) = r$. Assume by way of contradiction that $x \notin \mathcal{N}_(r)$, i.e., $x_r(t) > 0$ for some $t \in \mathrm{R}_+$. As $x - y \in \mathrm{L}_-$, we have $x_r - y_r \in \mathrm{L}_-$ (part (a) of Lemma 5). Thus, $y_r(t) \ge x_r(t) > 0$, which is a contradiction to $I_1(y) = r$. □

**Proof of Corollary 1.**

For any $H \in \mathrm{L}^*$ and $r \in \mathrm{R}$, denote by $\overline{H}_r$ the functional on $\mathrm{L}$ given by $x \mapsto H(x_r)$. Note that $\overline{H}_r \in \mathrm{L}^*$ (Lemma 3, part (d)); moreover, if $H \in \mathcal{NPV}$, then so does $\overline{H}_r$ for all $r \in \mathrm{R}_+$. Let the correspondence $\mathcal{N}: \mathrm{R}_+ \rightrightarrows \mathrm{L}$ be given by $\mathcal{N}(r) := \{x \in \mathrm{L} : x_r \in \mathrm{N}\}$.

(a)$\Rightarrow$(b). Since $\mathcal{N}$ is a cap, the correspondence $\mathcal{F}:\mathrm{R}_+ \rightrightarrows \mathcal{NPV}$ defined by $\mathcal{F}(r) := \mathcal{N}(r)^\circ \cap \mathcal{NPV}$ satisfies conditions 1–4 in part (c) of Proposition 4. Set $\Phi := \mathcal{F}(0)$. Properties 1' and 2' now follow from conditions 1 and 2. To establish property 3', pick $H \in \Phi$ and $r \in \mathrm{R}_{++}$. For any $x \in \mathcal{N}(0)$, we have $H(x) \le 0$ or, equivalently, $\overline{H}_r(x_{-r}) \le 0$. Since $x \mapsto x_{-r}$ maps $\mathcal{N}(0)$ onto $\mathcal{N}(r)$, we get $\overline{H}_r \in \mathcal{N}(r)^\circ$. Finally, as $\overline{H}_r \in \mathcal{NPV}$, we obtain $\overline{H}_r \in \mathcal{F}(r) \subseteq \mathcal{F}(0) = \Phi$, where the inclusion follows from condition 4. This proves property 3'.

(b)$\Rightarrow$(a). Define the correspondence $\mathcal{F}:\mathrm{R}_+ \rightrightarrows \mathcal{NPV}$ by $\mathcal{F}(r) := \{\overline{H}_r, H \in \Phi\}$ (note that $\overline{H}_r \in \mathcal{NPV}$, whenever $H \in \mathcal{NPV}$ and $r \in \mathrm{R}_+$; so $\mathcal{F}(r) \subseteq \mathcal{NPV}$). Let us show that $\mathcal{F}$ satisfies properties 1–4 in part (c) of Proposition 4.

Pick $r \in \mathrm{R}_+$. We have

$$\mathcal{N}(r) = \{x \in \mathrm{L} : x_r \in \mathrm{N}\} = \{x \in \mathrm{L} : x_r \in \Phi^\circ\} = \{x \in \mathrm{L} : H(x_r) \le 0 \ \forall H \in \Phi\}$$
$$= \{x \in \mathrm{L} : \overline{H}_r(x) \le 0 \ \forall H \in \Phi\} = \{\overline{H}_r, H \in \Phi\}^\circ = \mathcal{F}(r)^\circ,$$

where the second equality follows from condition 1'. So property 1 holds. Property 2 follows from condition 2' and the definition of $\mathcal{F}$. For any $r \ge s \ge 0$, we have $\mathcal{F}(r) = \{\overline{H}_r, H \in \Phi\} \subseteq \{\overline{H}_r, \overline{H}_{r-s} \in \Phi\} = \{\overline{H}_s, H \in \Phi\} = \mathcal{F}(s)$, where the inclusion follows from condition 3'. This proves property 4. Finally, to verify property 3, pick $H \in \mathcal{F}(r)$ and denote by $\alpha \in \mathcal{A}$ the discount function associated with $H$. By the definition of $\mathcal{F}(r)$, $\overline{H}_{-r} \in \Phi \subseteq \mathcal{NPV}$. So $t \mapsto e^{rt}\alpha(t)$ is a discount function; in particular, it is nonincreasing. Now property 3 follows from Lemma 6. □

**Proof of Lemma 2.**

Let $\mathcal{N}$ be a stable cap. For any $x \in \mathrm{L}$, let $h_x : \mathrm{R} \to \mathrm{L}$ be the map given by $h_x(r) := x_r$. Note that $h_x$ is continuous (Lemma 3, part (e)).

*Claim 1*: for all $r \in \mathrm{R}_+$, $\mathcal{N}(r) = \bigcap_{s:s>r} \mathcal{N}(s)$.

Choose $r \in \mathrm{R}_+$. By condition (iii) in the definition of a cap, $\mathcal{N}(r) \subseteq \bigcap_{s:s>r} \mathcal{N}(s)$. To prove the reverse inclusion, pick $x \in \bigcap_{s:s>r} \mathcal{N}(s)$ and note that $h_x((r,+\infty)) \subseteq \mathcal{N}(0)$ as $\mathcal{N}$ is stable. We have $h_x([r,+\infty)) = h_x(\mathrm{cl}((r,+\infty))) \subseteq \mathrm{cl}(h_x((r,+\infty))) \subseteq \mathrm{cl}(\mathcal{N}(0)) = \mathcal{N}(0)$, where the first inclusion follows from continuity of $h_x$ and the last equality comes from the fact that $\mathcal{N}(0)$ is closed (condition (vi) in the definition of a cap). Therefore, $x_r = h_x(r) \in \mathcal{N}(0)$, or equivalently, $x \in \mathcal{N}(r)$. This proves that $\mathcal{N}(r) = \bigcap_{s:s>r} \mathcal{N}(s)$.

*Claim 2*: for all $r \in \mathrm{R}_{++}$, $\mathcal{N}(r) = \mathrm{cl}\left( \bigcup_{s:s<r} \mathcal{N}(s) \right)$.

Choose $r \in \mathrm{R}_{++}$. By condition (iii) in the definition of a cap, $\bigcup_{s:s<r} \mathcal{N}(s) \subseteq \mathcal{N}(r)$. Therefore, $\mathrm{cl}\left(\bigcup_{s:s<r} \mathcal{N}(s)\right) \subseteq \mathrm{cl}(\mathcal{N}(r)) = \mathcal{N}(r)$, where the last equality follows from condition (vi) in the definition of a cap. To prove the reverse inclusion, pick $x \in \mathcal{N}(r)$ and let $\mathrm{U}$ be a neighborhood of $x$. Since $h_x$ is continuous, there is $\varepsilon \in (0,r]$ such that $x_\varepsilon = h_x(\varepsilon) \in \mathrm{U}$. As $x \in \mathcal{N}(r)$ and $\mathcal{N}$ is stable, we have $x_\varepsilon \in \mathcal{N}(r-\varepsilon) \subseteq \bigcup_{s:s<r} \mathcal{N}(s)$. Therefore, $\mathrm{U}$ intersects $\bigcup_{s:s<r} \mathcal{N}(s)$. This proves that $\mathcal{N}(r) \subseteq \mathrm{cl}\left(\bigcup_{s:s<r} \mathcal{N}(s)\right)$. □

**Proof of Proposition 5.**

(a)$\Rightarrow$(b). Let $\mathcal{N}$ be a continuous $E$-cap and let $\overline{E}: \mathrm{L} \to \mathrm{R}_+ \cup \{+\infty\}$ be defined by $\overline{E}(x) := \inf\{r \in \mathrm{R}_+ : x \in \mathcal{N}(r)\}$ (with the convention $\inf \varnothing = +\infty$). We shall show that $\overline{E}$ is a refinement of $E$ and $\mathcal{N}(r) = \{x \in \mathrm{L} : \overline{E}(x) \le r\}$ $\forall r \in \mathrm{R}_+$.

Properties (I), (II), and (III) follow, respectively, from conditions (i), (ii), and (v) in the definition of an $E$-cap.

Pick $r \in \mathrm{R}_+$. It follows from the definition of $\overline{E}$ and condition (iii) that $\overline{E}(x) \le r$ $\Rightarrow$ $x \in \mathcal{N}(r+\varepsilon)$ $\forall \varepsilon > 0$. Since the cap $\mathcal{N}$ is continuous, the latter condition implies $x \in \mathcal{N}(r)$. On the other hand, by construction, $x \in \mathcal{N}(r)$ $\Rightarrow$ $\overline{E}(x) \le r$. This proves that $\{x \in \mathrm{L} : \overline{E}(x) \le r\} = \mathcal{N}(r)$. Property (IV) now follows from conditions (iv)–(vi).

Finally, to verify property (V), choose $r \in \mathrm{R}_{++}$ and note that $\{x \in \mathrm{L} : \overline{E}(x) < r\} = \bigcup_{s:s<r} \{x \in \mathrm{L} : \overline{E}(x) \le s\}$. We have

$$\mathrm{cl}(\{x \in \mathrm{L} : \overline{E}(x) < r\}) = \mathrm{cl}\left(\bigcup_{s:s<r} \{x \in \mathrm{L} : \overline{E}(x) \le s\}\right) = \mathrm{cl}\left(\bigcup_{s:s<r} \mathcal{N}(s)\right) = \mathcal{N}(r) = \{x \in \mathrm{L} : \overline{E}(x) \le r\},$$

where the third equality follows from continuity of the cap $\mathcal{N}$.

(b)$\Rightarrow$(a). Assume that $\overline{E}$ is a refinement of $E$ and $\mathcal{N}(r) = \{x \in \mathrm{L} : \overline{E}(x) \le r\}$ $\forall r \in \mathrm{R}_+$. We must show that $\mathcal{N}$ is a continuous $E$-cap. Properties (i) and (ii) in the definition of an $E$-cap follow, respectively, from conditions (I) and (II). Property (iii) holds trivially. Conditions (III) and (IV) imply properties (iv)–(vi). Finally, we show that the cap $\mathcal{N}$ is continuous. The equality $\mathcal{N}(r) = \bigcap_{s:s>r} \mathcal{N}(s)$, $r \in \mathrm{R}_+$ holds trivially, whereas the equality $\mathcal{N}(r) = \mathrm{cl}\left(\bigcup_{s:s<r} \mathcal{N}(s)\right)$, $r \in \mathrm{R}_{++}$ follows from the identity $\{x \in \mathrm{L} : \overline{E}(x) < r\} = \bigcup_{s:s<r} \{x \in \mathrm{L} : \overline{E}(x) \le s\}$ and condition (V). □

**Proof of Proposition 8.**

(a)$\Rightarrow$(b). The proof is similar to that of "(b)$\Rightarrow$(c)" in Proposition 4. We omit the details.

(b)$\Rightarrow$(a). Pick $r \in \mathrm{R}_+$. Conditions (ii)', (iii)', (iv)–(vi) in the definition of an $E$-floor follow from conditions 1 and 4 in part (b). To verify condition (i)', choose $x \in \mathrm{P}$ and set $s = E(x) = I_1(x)$. First, assume that $s < r$. As $I_2(x) = I_1(x) < r$, we have $F_r(x) < 0$ by the definition of $I_2(x)$. Conditions 1 and 2 in part (b) imply $x \notin \mathcal{N}(r)$. Now assume that $s \geq r$. Pick $F^{(\alpha)} \in \mathcal{F}(r)$ and set $\beta(t) := e^{st}\alpha(t)$. From condition 3 in part (b) it follows that $\beta$ is nondecreasing. $\beta$ is continuous (if $\beta$ was nondecreasing and discontinuous, then $\alpha$ would not belong to $\mathcal{A}$). Let $T$ be the maturity date of $x$. Using integration by substitution and integration by parts, we get

$$F^{(\alpha)}(x) = x(0) + \int_0^T \alpha(t)\mathrm{d}x(t) = x(0) + \int_0^T e^{st}\alpha(t)\mathrm{d}\big(x_s(t) - x(0)\big) = x(0) + \int_0^T \beta(t)\mathrm{d}x_s(t)$$

$$= x(0) + \beta(T)x_s(T) - \beta(0)x_s(0) - \int_0^T x_s(t)\mathrm{d}\beta(t) = -\int_0^T x_s(t)\mathrm{d}\beta(t) \geq 0\,,$$

where the last equality follows from the equalities $\beta(0) = 1$, $x_s(0) = x(0)$, and $x_s(T) = x_s(+\infty) = 0$ (by the definition of $I_1(x)$), whereas the inequality follows from the facts that $\beta$ is nondecreasing and $x_s$ is nonpositive. Since $F^{(\alpha)}$ was an arbitrary element of $\mathcal{F}(r)$, condition 1 in part (b) implies that $x \in \mathcal{N}(r)$. □

**Lemma 7.**

*For $\alpha \in \mathcal{A}$ and $r \in \mathrm{R}_+$, the following conditions are equivalent.*

(a) *The function $t \mapsto e^{rt}\alpha(t)$ is nondecreasing.*

(b) *$\alpha$ is positive, $\ln\alpha$ is absolutely continuous, and $-(\ln\alpha(t))' \leq r$, whenever the derivative in the left-hand side of the inequality exists.*

**Proof.**

(a)$\Rightarrow$(b). Condition (a) implies that $\alpha$ is positive. For any $0 \leq s \leq t$, we have

$$0 \leq \ln\alpha(s) - \ln\alpha(t) \leq r(t-s)\,, \tag{9}$$

where the first inequality holds by virtue of $\alpha \in \mathcal{A}$ and the second inequality follows from (a). This proves that $\ln\alpha$ is Lipschitz continuous (with the Lipschitz constant $r$) and, hence, absolutely continuous. The second part of (9) implies that $-(\ln\alpha(t))' \leq r$, whenever the derivative in the left-hand side of the inequality exists.

(b)$\Rightarrow$(a). Pick $0 \leq s < t$. Since $\ln\alpha$ is absolutely continuous, we have $\ln\alpha(s) - \ln\alpha(t) = \int_s^t (-\ln\alpha(\tau))'\mathrm{d}\tau \leq \int_s^t r\mathrm{d}\tau = r(t-s)$. Rearranging the terms and taking the exponential of both sides, we get $e^{rs}\alpha(s) \leq e^{rt}\alpha(t)$. □

## 9. References

Alchian, A.A.: The rate of interest, Fisher's rate of return over costs and Keynes' internal rate of return. Am. Econ. Rev. **45**(5), 938–943 (1955)

Aliprantis, C.D., Border, K.C.: Infinite dimensional analysis: a hitchhiker's guide. Springer, Berlin (2006)

Armerin, F.: An axiomatic approach to the valuation of cash flows. Scand. Actuar. J. **2014**(1), 32–40 (2014)

Arrow, K.J., Levhari, D.: Uniqueness of the internal rate of return with variable life of investment. Econ. J. **79**(315), 560–566 (1969)

Basu, S., Pollack, R., Roy, M.-F.: Algorithms in real algebraic geometry. Springer-Verlag, Berlin (2003)

Beaves, R.G.: Net present value and rate of return: implicit and explicit reinvestment rate assumption. Eng. Econ. **33**(4), 275–302 (1988)

Bidard, C.: Fixed capital and internal rate of return. J. Math. Econ. **31**, 523–541 (1999)

Bronshtein, E.M.: Limitary profitability of financial operations [Paper presentation]. The 17th International AFIR Colloquium, Stockholm, Sweden (2007, June). https://actuaries.org/app/uploads/2025/10/Bronshtein_Stockholm2007.pdf

Bronshtein, E.M., Skotnikov D.A.: The limit profitability of financial operations. J. Appl. Ind. Math. **1**(2), 165–174 (2007)

Calice, P., Kalan, F.D., Masetti, O.: Interest rate repression around the world. EFI Insight-Finance. The World Bank (2020)

Cantor, D.G., Lippman, S.A.: Investment selection with imperfect capital markets. Econometrica **51**(4), 1121–1144 (1983)

Dudley, C.L.Jr.: A note on reinvestment assumptions in choosing between net present value and internal rate of return. J. Finance. **27**(4), 907–915 (1972)

Ferrari, A., Masetti, O., Ren, J.: Interest rate caps: the theory and the practice. Policy Research Working Paper Series. No. 8398. The World Bank (2018)

Gronchi, S.: On investment criteria based on the internal rate of return. Oxf. Econ. Pap. **38**(1), 174–180 (1986)

Hartman, J.C., Schafrick, I.C.: The relevant internal rate of return. Eng. Econ. **49**(2), 139–158 (2004)

Hazen, G.B.: A new perspective on multiple internal rates of return. Eng. Econ. **48**(1), 31–51 (2003)

Herbst, A.: The unique, real internal rate of return: caveat emptor! J. Financ. Quant. Anal. **13**(2), 363–370 (1978)

Hungerbühler, N., Wasem, M.: An integral that counts the zeros of a function. Open Math. **16**, 1621–1633 (2018)

Jameson, G.: Ordered linear spaces. Springer-Verlag (1970)

Kadets, V.: A course in functional analysis and measure theory. Springer (2018)

Karp, I.M.: An actuary’s perspective on criminal rate of interest calculations. Canadian Institute of Actuaries (2003)

Keane, S.M.: The internal rate of return and the reinvestment fallacy. Abacus **15**(1), 48–55 (1979)

Krantz, S.G., Parks, H.G.: A primer on real analytic functions. Birkhäuser (2002)

Lin, A.Y.S.: The modified internal rate of return and investment criterion. Eng. Econ. **21**(4), 237–247 (1976)

Magni, C.A.: Average internal rate of return and investment decisions: a new perspective. Eng. Econ. **55**(2), 150–181 (2010)

Magni, C.A.: Capital depreciation and the underdetermination of rate of return: a unifying perspective. J. Math. Econ. **67**, 54–79 (2016)

Magni, C.A., Martin, J.D.: The two sides of the reinvestment assumption fallacy in IRR and NPV. Eng. Econ. **70**(3), 122−151 (2025)

Maimbo, S.M., Gallegos, C.A.H.: Interest rate caps around the world: still popular, but a blunt instrument. Policy Research Working Paper Series. No. 7070. The World Bank (2014)

Miller, Jr.T.W., Black, H.A.: Examining arguments made by interest rate cap advocates. In Peirce H., Klutsey, B. (eds.) Reframing financial regulation: enhancing stability and protecting consumers. George Mason University (2016)

Monteiro, G.A., Slavík, A., Tvrdý, M.: Kurzweil-Stieltjes integral. Theory and applications. World Scientific (2018)

Narici, L., Beckenstein, E.: Topological vector spaces. CRC Press (2010)

Norberg, R.: Payment measures, interest, and discounting: an axiomatic approach with applications to insurance. Scand. Actuar. J. **1**, 14–33 (1990)

Norstrøm, C.J.: A sufficient condition for a unique nonnegative internal rate of return. J. Financ. Quant. Anal. **7**(3), 1835–1839 (1972)

Promislow, S.D.: Axioms for the valuation of payment streams: a topological vector space approach. Scand. Actuar. J. **2**, 151–160 (1994)

Promislow, S.D.: Classification of usurious loans. In: Sherris, M. (ed.) Proceedings of the 7th International AFIR Colloquium, pp. 739–759. Institute of Actuaries of Australia, Sydney (1997)

Promislow, S.D.: Fundamentals of actuarial mathematics. Wiley (2015)

Promislow, S.D., Spring, D.: Postulates for the internal rate of return of an investment project. J. Math. Econ. **26**, 325–361 (1996)

Rich, S.P., Rose, J.T.: Re-examining an old question: does the IRR method implicitly assume a reinvestment rate? J. Financ. Educ. **40**(1/2), 152–166 (2014)

Shull, D.M.: Efficient capital project selection through a yield-based capital budgeting technique. Eng. Econ. **38**(1), 1–18 (1992)

Sokolov, M.V.: NPV, IRR, PI, PP, and DPP: a unified view. J. Math. Econ. **114**, 102992 (2024)

Solomon, E.: The arithmetic of capital-budgeting decisions. J. Bus. **29**(2), 124–129 (1956)

Spring, D.: General balance functions in the theory of interest. arXiv preprint arXiv:1208.1479 (2012)

Stiglitz, J.E., Weiss, A.: Credit rationing in markets with imperfect information. Am. Econ. Rev. **71**(3), 393–410 (1981)

Teichroew, D., Robichek, A.A., Montalbano, M.: An analysis of criteria for investment and financial decisions under certainty. Manag. Sci. **12**(3), 151–179 (1965)

Uniform Law Conference of Canada: Section 347 of the Criminal Code in need of reform. Report of the criminal section working group on criminal interest rate: a discussion paper (2008)

Vilensky, P.L., Smolyak, S.A.: Internal rate of return of project and its modifications. Audit and Financial Analysis **4**, 203–225 (1999)

Waldron, M.A.: Section 347 of the Criminal Code: A deeply problematic law. Uniform Law Conference (2002)

Weber, T.A.: On the (non-)equivalence of IRR and NPV. J. Math. Econ. **52**, 25–39 (2014)

Widder, D.V.: The Laplace transform. Princeton University Press (1946)